\documentclass[lettersize,journal]{IEEEtran}%[sigconf, 9pt]{acmart}
\IEEEoverridecommandlockouts
\usepackage{geometry}
\pdfoutput=1

\usepackage{url}
\usepackage{amsmath,amsfonts}
\usepackage{amssymb}
\usepackage{algorithm}
\usepackage{booktabs}    
\usepackage{subcaption}

\usepackage{algpseudocode}
\usepackage[]{graphicx}
\graphicspath{ {./figs/} }
\usepackage{textcomp}
\usepackage{xcolor}

\usepackage{comment}
\usepackage{multirow}
\usepackage{epsfig}
\usepackage{etoolbox}
\usepackage{array}
\usepackage{tabularx}
\usepackage{hhline}
\usepackage{xspace}
\usepackage{upgreek}
\usepackage{makecell}

\def\BibTeX{{\rm B\kern-.05em{\sc i\kern-.025em b}\kern-.08em
    T\kern-.1667em\lower.7ex\hbox{E}\kern-.125emX}}

\newcounter{itemlistc}

\renewcommand{\baselinestretch}{1.}

\graphicspath{{./figs/}}
\begin{document}

% ================================= Title =================================================================
%\title{{BPINN-EM-Post}: \underline{B}ayesian \underline{P}hysics-\underline{I}nformed \underline{N}eural \underline{N}etwork based Stochastic \underline{E}lectro\underline{m}igration Damage Analysis in the \underline{Post}-void Phase}

% Original title
\title{Accelerating Physics-Based Electromigration Analysis via Rational Krylov Subspaces}
% ================================= Title =================================================================

\author{
\IEEEauthorblockN{Sheldon X.-D. Tan,~\IEEEmembership{Fellow,~IEEE} and Haotian Lu,~\IEEEmembership{Student Member,~IEEE}}    \\     
\IEEEauthorblockA{\textit{Department of Electrical and Computer Engineering}, \textit{University of California at Riverside}, \textit{Riverside, CA}} 
%\small \textit{stan,hlu123}\@ece.ucr.edu, 
%Department of Electrical and Computer Engineering, University of California\\
% \small \textit{\{slami002, hlu123\}@ucr.edu, stan@ece.ucr.edu}
\thanks{This work is supported in part by NSF grants under No. CCF-2007135, and in part by NSF grant under No. CCF-2305437.}
}

\maketitle

\begin{abstract}
Electromigration (EM)-induced stress in interconnect wires remains one of the key design challenges for nanometer-scale VLSI designs. 
% For fast evaluation of EM damage, one has to solve partial differential equation of stress evolution in the confined metal interconnect trees, which is a time consuming process.  
To enable fast evaluation of EM-induced damage, it is necessary to solve the partial differential equations (PDEs) governing stress evolution in confined metal interconnect trees, which is a time-consuming process.
In this work, we propose two fast EM stress analysis techniques based on {\it rational Krylov subspace} reduction. 
% Unlike traditional Krylov subspace method, which can be viewed as expansion around infinite time (at frequency of 0), rational Krylov subspace allows expansion at specific time constant, which can used to be aligned with application metric such as nucleation time in our application. 
Unlike traditional Krylov subspace methods, which effectively expand around zero frequency (infinite time), rational Krylov subspaces allow expansion at specific time constants, enabling better alignment with application-specific metrics such as nucleation time.
As a result, highly compact models can be achieved with negligible accuracy loss.  
We explore rational Krylov subspace reduction through two simulation frameworks: one in the frequency domain via the extended rational Krylov subspace method, termed {\it ExtRaKrylovEM}, and one in the time domain via the rational Krylov exponential integration (EI) method, termed {\it EiRaKrylovEM}. 
We show that the accuracy of both methods is sensitive to the choice of expansion point, or equivalently, shift time.
We further show that an effective shift time is typically close to a time of interest, such as the nucleation time or the steady-state time, which provides a practical starting point for determining the optimal shift time at a given reduction order.   
To further improve fidelity, we develop a coordinate descent optimization method that identifies the optimal reduction order and shift times for both nucleation and post-void phases. 
Experimental results on synthesized structures and industrial benchmarks demonstrate that {\it ExtRaKrylovEM} and {\it EiRaKrylovEM} achieve orders-of-magnitude improvements over the standard finite-difference method in both efficiency and accuracy. 
Specifically, using only 4--6 reduction orders, our methods deliver sub-0.1\% error in nucleation-time and resistance-change predictions while providing 20--500$\times$ speedup over finite-difference solutions, depending on the number of simulation steps. 
In stark contrast, standard extended Krylov methods require 50+ orders yet still exhibit 10--20\% nucleation time errors, rendering them impractical for EM-aware optimization and stochastic EM analysis.

% Numerical results demonstrate that {\it BPINN-EM-Post} achieves over $240 \times$ speedup compared to Monte Carlo simulations using the FEM-based COMSOL solver and more than $65 \times$  speedup compared to Monte Carlo simulations using the FDM-based EMSpice method.
%Numerical results demonstrate that {\it %\textbf{ExtRaKrylovEM:}} 
\end{abstract}


% \vspace{-5pt}
\section{Introduction}
\label{sec:introduction}

% ===================================== Clean Version ===============================================
Electromigration (EM) is a key reliability concern in VLSI interconnects, arising from momentum transfer between high-current-carrying electrons and metal atoms. 
This atomic flux induces compressive stress near the anode and tensile stress near the cathode, generating a back stress that counteracts the atomic flux. When the stress exceeds a critical threshold, voids or hillocks can form to relax the stress, leading to open or short failures. 

Electromigration was traditionally modeled using empirical formulas such as Black's equation~\cite{Black:1969fc} and Blech's product~\cite{Blech:1976ko}, which relate the mean time to failure (MTTF) of a wire segment to its current density and temperature. However, 
these empirical models have faced growing criticism for being overly conservative and applicable only to a single wire segment~\cite{Hauschildt:2013cv,Sukharev:2013tq}. To address these issues, several 
physics-based EM models and assessment techniques based on Korhonen's equations~\cite{Korhonen:jap1993} 
have recently been proposed~\cite{deOrio:2010,HuangTan:TCAD'16,sukharev2016postvoiding,ChenTan:TCAD'16,
MishraSapatnekar:2016DAC,ChenTan:TDMR'17,
Chatterjee:2018TCAD,CookSun:TVLSI'18,WangSun:ICCAD'17,ZhaoTan:TVLSI'18,
Abbasinasab:DAC'2018,ChenTan:TVLSI'19,SunYu:TDMR'20,Shohel:ICCAD'21,TanTahoori:Book'19,Shohel:ICCAD'21,Stoikos:SMACD2021}.

Unlike traditional EM models that rely on current density in a single wire segment, physics-based EM models consider stress evolution across multiple wire segments in an interconnect tree, capturing complex interactions and correlations among them~\cite{HauRiege:2001JAP}. At the core of these approaches lies the numerical solution of Korhonen's partial differential equation (PDE) for hydrostatic stress evolution under blocking boundary conditions~\cite{Korhonen:jap1993}. While analytical approximations exist for limited structures~\cite{ChenTan:TDMR'17,wang:TCAD'21}, general tree topologies require numerical methods such as finite-difference discretization, which become computationally expensive for large designs. To evaluate stochastic reliability metrics such as MTTF, Monte Carlo (MC) analysis is typically employed~\cite{Chatterjee:ICCAD'16,Najm:IRPS2019}, further amplifying the computational burden.

In this work, we address the computational challenges of physics-based EM analysis by accelerating the numerical solution of Korhonen's equation through a novel {\it rational Krylov subspace scheme}. Unlike traditional Krylov methods that expand around zero frequency (infinite time), the rational Krylov subspace method allows expansion at specific time constants, enabling alignment with application metrics such as nucleation time in our context. This leads to highly compact models with minimal error. We exploit the rational Krylov subspace concept in two complementary simulation frameworks: in the frequency domain through the extended rational Krylov subspace method, and in the time domain through the rational Krylov exponential integration (EI) method. Our major contributions are as follows:

\begin{itemize}
\item First, we propose a novel \emph{extended rational Krylov subspace} method, called \textit{ExtRaKrylovEM}, that integrates the extended Krylov framework with the rational subspace concept to efficiently reduce the EM matrices derived from Korhonen's equation in the frequency domain. The rational Krylov approach shifts the projection subspace toward critical time regions of interest in the stress evolution, such as the nucleation phase. This capability enables significantly higher accuracy with much smaller reduction orders compared to conventional extended Krylov methods. \textit{ExtRaKrylovEM} also naturally mitigates the singularity issue in the EM matrix inversion process.

\item Second, we extend the rational Krylov subspace concept to the time domain through an exponential integration (EI) framework, yielding the \textit{EiRaKrylovEM} method. This approach constructs a rational Krylov subspace to efficiently approximate the matrix exponential action on the initial stress vector, enabling accurate time-domain simulation of stress evolution with substantially reduced computational cost. By strategically selecting shift times aligned with critical phases of stress evolution, \textit{EiRaKrylovEM} achieves significant speedups over conventional time-stepping methods while maintaining high accuracy.

\item To determine the optimal reduction order and shift parameters for both nucleation and post-void phases, we develop a coordinate-descent optimization algorithm driven by application-specific metrics, namely nucleation time (determined by critical stress thresholds) and resistance change in voided segments (also stress-dependent). This contrasts with existing rational Krylov methods that rely on generic frequency- or time-domain error metrics defined over all system dynamics. By directly optimizing for EM-relevant performance metrics, our approach yields reduced-order models that more accurately capture the critical stress dynamics governing interconnect reliability.

%\item The resulting {\it ExtRaKrylovEM} method, typically requires only a reduction order of 4--5 to achieve less than 1\% error in predicting nucleation time and resistance change (when nucleation occurs), compared to the 20--30 order commonly needed by conventional Krylov-based methods for the same accuracy. This efficiency makes {\it ExtRaKrylovEM} particularly advantageous for Monte Carlo (MC) analyses, where both computational speed and numerical accuracy are critical for reliable stochastic evaluation.

\item Experimental results on a set of synthesized structures and industrial benchmarks (one ARM design and several IBM benchmarks) show that the proposed {\it ExtRaKrylovEM} and {\it EiRaKrylovEM} achieve orders-of-magnitude improvements in both efficiency and accuracy. Specifically, using only 4--6 reduction orders, our methods deliver sub-0.1\% error in nucleation-time and resistance-change predictions while providing 20--500$\times$ speedup over finite-difference solutions, depending on the number of simulation time steps. In stark contrast, standard extended Krylov methods require 50+ orders yet still exhibit 10--20\% nucleation-time errors, rendering them impractical for EM-aware optimization and stochastic EM analysis. 

%The computational efficiency of our approach enables full-chip Monte Carlo analysis that was previously infeasible, opening new pathways for realistic reliability assessment of advanced VLSI interconnects.

\end{itemize}

The paper is organized as follows. In Section~\ref{sec:review}, we review related work on EM analysis and Krylov subspace methods. In Section~\ref{sec:prelim}, we present the preliminaries of EM stress-evolution modeling and introduce rational Krylov subspace methods. In Section~\ref{sec:extrakrylov_method}, we detail the proposed {\it ExtRaKrylovEM} method and present the coordinate-descent algorithm. In Section~\ref{sec:rakrylov_ei_method}, we describe the {\it EiRaKrylovEM} method. In Section~\ref{sec:results}, we present experimental results. Finally, we conclude the paper in Section~\ref{sec:Conclusion}.

\section{Related Work}
\label{sec:review}

Krylov subspace methods have been widely used for model-order reduction in various domains~\cite{Antoulas:Book'2005}, including VLSI circuit simulation~\cite{Tan:Book'07}. Existing methods mainly target large dynamic systems, but they can be inefficient when the numbers of inputs and outputs are large. To mitigate this issue, the extended Krylov subspace method~\cite{Wang:DAC'00} and Krylov-based exponential integration (EI) frameworks~\cite{Bochev:short_guide2012} have been proposed, both of which naturally incorporate input excitations into the reduced model. A second limitation of traditional Krylov methods is that they typically expand the subspace around zero frequency (infinite time), which may not effectively capture the dynamics of interest in EM stress evolution.

\begin{figure}[h]
  \vspace{-5pt}
  \centering
  \includegraphics[width=0.85\columnwidth]{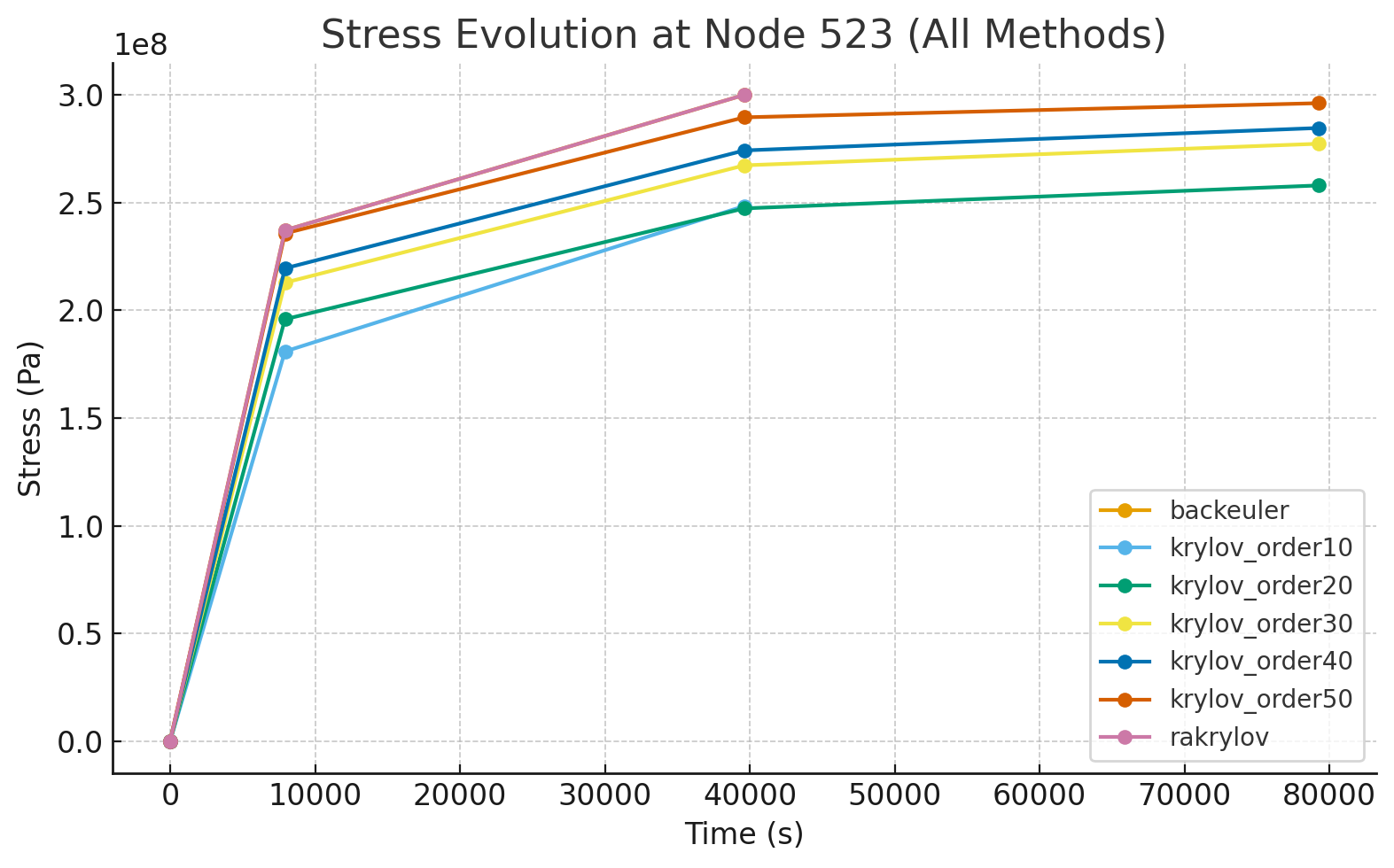}
  \vspace{-5pt}
  \caption{\small Nucleation times at different reduction orders using the standard Krylov subspace method for a 100-segment wire (critical stress $= 10^8$ Pa)}
  \label{fig:stress_diff_orders_wire0_seg100}
  \vspace{-5pt}
\end{figure}

For EM analysis, Krylov subspace methods have been applied in \emph{FastEM}, where the extended Krylov subspace framework is implemented within a finite-difference formulation~\cite{CookSun:TVLSI'18}. The standard Krylov approach can be interpreted as a subspace expansion around the frequency point $s = 0$ (corresponding to time $t \to \infty$). Consequently, it often requires a high reduction order to accurately capture transient behavior such as the nucleation phase, which occurs far from the asymptotic regime. As shown in Fig.~\ref{fig:stress_diff_orders_wire0_seg100},
the standard Krylov method exhibits a significant deviation in nucleation-time prediction (critical stress of $10^8$~Pa) even at order~50: the estimated
nucleation time is about $8\times10^{4}$~s, compared with $4\times10^{4}$~s obtained by the Backward-Euler reference. Moreover, the numerical error may
lead to incorrect identification of the cathode nucleation node. In contrast, when the extended Krylov subspace method uses a shift time aligned
with the nucleation time and an order of 5 (e.g., $t_{\mathrm{shift}} = 3.67\times10^{4}$~s), the results match the Backward-Euler solution almost exactly.

Rational Krylov subspaces have been applied in a variety of numerical modeling contexts, including the solution of Maxwell’s equations~\cite{Botchev:JCA2016}, VLSI power-grid analysis~\cite{Zhuang:TCAD2016}, and, more recently, EM stress
analysis~\cite{Stoikos:SMACD23}, all within the exponential integration (EI) framework. In the EI method, the Krylov subspace is used to approximate the matrix-exponential operation $e^{tA}\nu$ in the time domain, where $t$ is the time variable and $\nu$ is the initial state vector. A fundamental limitation of this approach is that time-varying input sources require recomputation of the Krylov subspace whenever the input changes, because the computed subspace only incorporates the input excitation from the previous time step. This makes EI methods inefficient for many practical analyses.

In the EM formulation of~\cite{Stoikos:SMACD23}, a rational Krylov method is also employed; however, it constructs the Krylov subspace using $(A^{-1} - s_0 I)^{-1}$ rather than $(A - s_0 I)^{-1}$ as adopted in~\cite{Zhuang:TCAD2016} and in our work, where $A$ is the EM operator defined in Eq.~\eqref{eq:lti_em}. This choice biases the Krylov basis toward the dominant eigenvalues of $A$, thereby emphasizing high-frequency (short-transient) dynamics. In electromigration analysis, however, the primary quantities of interest, namely nucleation time and post-void stress evolution, are governed by the smallest eigenvalues of $A$, which correspond to long-tail, near-steady-state behavior. As shown in~\cite{Zhuang:TCAD2016}, using $A^{-1}$ (or its shifted
variant) is preferable because the largest eigenvalues of $A^{-1}$ correspond to the smallest eigenvalues of $A$, precisely the spectral components that dominate EM-induced failure mechanisms. 

Furthermore, the method in~\cite{Stoikos:SMACD23} selects the shift time or expansion point and reduction order purely based on minimizing the global model residual error, rather than application-specific performance metrics such as nucleation time or resistance evolution. This often leads to unnecessarily high reduction orders that emphasize short transient effects rather than the long-tail dynamics relevant to EM-induced failure analysis.
% Moreover, the methodology in~\cite{Stoikos:SMACD23} again assumes constant current excitation, which is insufficient for accurate EM modeling under realistic, time-varying operating conditions~\cite{Sukharev:JAP2015}.

In this work, we exploit the rational Krylov subspace method in two simulation frameworks: one in the frequency domain via the extended rational Krylov subspace method, and one in the time domain via the rational Krylov exponential integration (EI) method. In the frequency-domain approach, we apply the rational Krylov subspace method to approximate the EM dynamic system directly in the frequency domain within a model-order reduction framework that naturally incorporates arbitrary input-current waveforms through the extended Krylov subspace. This formulation avoids the matrix-inversion singularity issues encountered in traditional finite-difference methods during the nucleation phase~\cite{CookSun:TVLSI'18}. In the time-domain approach, we develop a rational Krylov EI method that constructs the Krylov subspace using $(A - s_0 I)^{-1}$, enabling efficient approximation of the matrix exponential action on the initial stress vector while focusing on the low-frequency dynamics relevant to EM reliability assessment.

% Also the selection of expansion points or shift parameters in this approach is driven by global model residual criteria rather than by application-specific performance metrics such as nucleation time or resistance evolution. This often leads to unnecessarily high reduction orders
% that emphasize short transient effects rather than the long-tail dynamics relevant to EM-induced failure analysis.

\section{Preliminaries}
\label{sec:prelim}

\subsection{EM Stress Modeling in Brief}
In confined metal wires subjected to high current densities, EM occurs as atoms are migrated from the cathode to the anode due to interactions between electrons and metal 
atoms~\cite{Black:1969fc}. 
Over time, this atom migration can cause void and hillock formation that compromises interconnect functionality. Existing current-density-based models such as Blech's limit~\cite{Blech:1976ko} and Black's MTTF~\cite{Black:1969fc} consider only one wire segment at a time and are widely regarded as overly conservative. 

Recently, physics-based EM modeling has been proposed, which reduces to solving Korhonen's equation for multi-segment interconnects in the same metallization tree~\cite{Korhonen:1993bb}. The equation for the nucleation phase in multi-segment interconnects is given by Eq.~\eqref{eq:korhonen_multisegment_nucleation},
\begin{equation}
\small
    \begin{aligned}
        & PDE: \frac{\partial \sigma_{ij}(x,t)}{\partial t}=\frac{\partial
        }{\partial x}\left[\kappa_{ij}(\frac{\partial \sigma_{ij}(x,t)}{\partial x} + G_{ij})\right],\ t>0  \\
        & BC: \sigma_{ij_1}(x_i,t)=\sigma_{ij_2}(x_i,t),\ t>0 \\
        & BC_{neu}: \sum_{ij} \kappa_{ij}(\frac{\partial \sigma_{ij}(x,t)}{\partial x} \big |_{x=x_{junc}}+G_{ij}) = 0,\ t>0 \\
        & BC_{neu}: \kappa_{ij}(\frac{\partial \sigma_{ij}(x,t)}{\partial x} \big |_{x=x_b}+G_{ij}) = 0,\ t>0 \\
        & IC: \sigma_{ij}(x,0)= \sigma_{ij,T}
        \label{eq:korhonen_multisegment_nucleation}
    \end{aligned}
\end{equation}
here $\sigma_{ij}(x,t)$ represents the stress in the interconnect segment $ij$ connecting nodes $i$ and $j$. 
In Eq.~\eqref{eq:korhonen_multisegment_nucleation}, $G_{ij}$ denotes the EM driving force in segment $ij$, calculated as $G_{ij} = \frac{e \rho J_{ij}Z^*}{\Omega}$, where $J_{ij}$ is the current density in segment $ij$. 
The stress diffusivity $\kappa_{ij}$ is defined by $\kappa_{ij} = D_a B\Omega/(k_B T)$, where $D_a$ is the effective atomic diffusion coefficient, $B$ is the effective bulk modulus, 
$k_B$ is Boltzmann's constant, and $T$ is the absolute temperature.
$e$ is the electron charge, $\rho$ is resistivity, and $Z^*$ is the effective charge.

The first BC in Eq.~\eqref{eq:korhonen_multisegment_nucleation} enforces stress continuity at inter-segment junctions, specifically at $x = x_{junc}$.
The second BC is a Neumann condition that enforces atomic-flux conservation at inter-segment junctions $x_{junc}$, while the third Neumann BC applies at blocking terminal boundaries $x = x_b$ to ensure zero atomic flux. 
The IC specifies the initial stress distribution in segment $ij$, denoted by $\sigma_{ij,T}$, which is typically zero.

In multi-segment interconnect trees, void nucleation occurs at the cathode when the steady-state nucleation stress surpasses the critical stress $\sigma_{\mathrm{crit}}$.
This moment marks the nucleation time, denoted as $t_{nuc}$. 
Beyond this time, $t > t_{nuc}$, the void grows larger and may lead to wire segment resistance changes. Once nucleated, the stress at the interface of the void dramatically drops to zero~\cite{Sukharev:tdmr'16}. 

\begin{figure}[h]
  \vspace{-5pt}
  \centering
  \includegraphics[width=0.99\columnwidth]{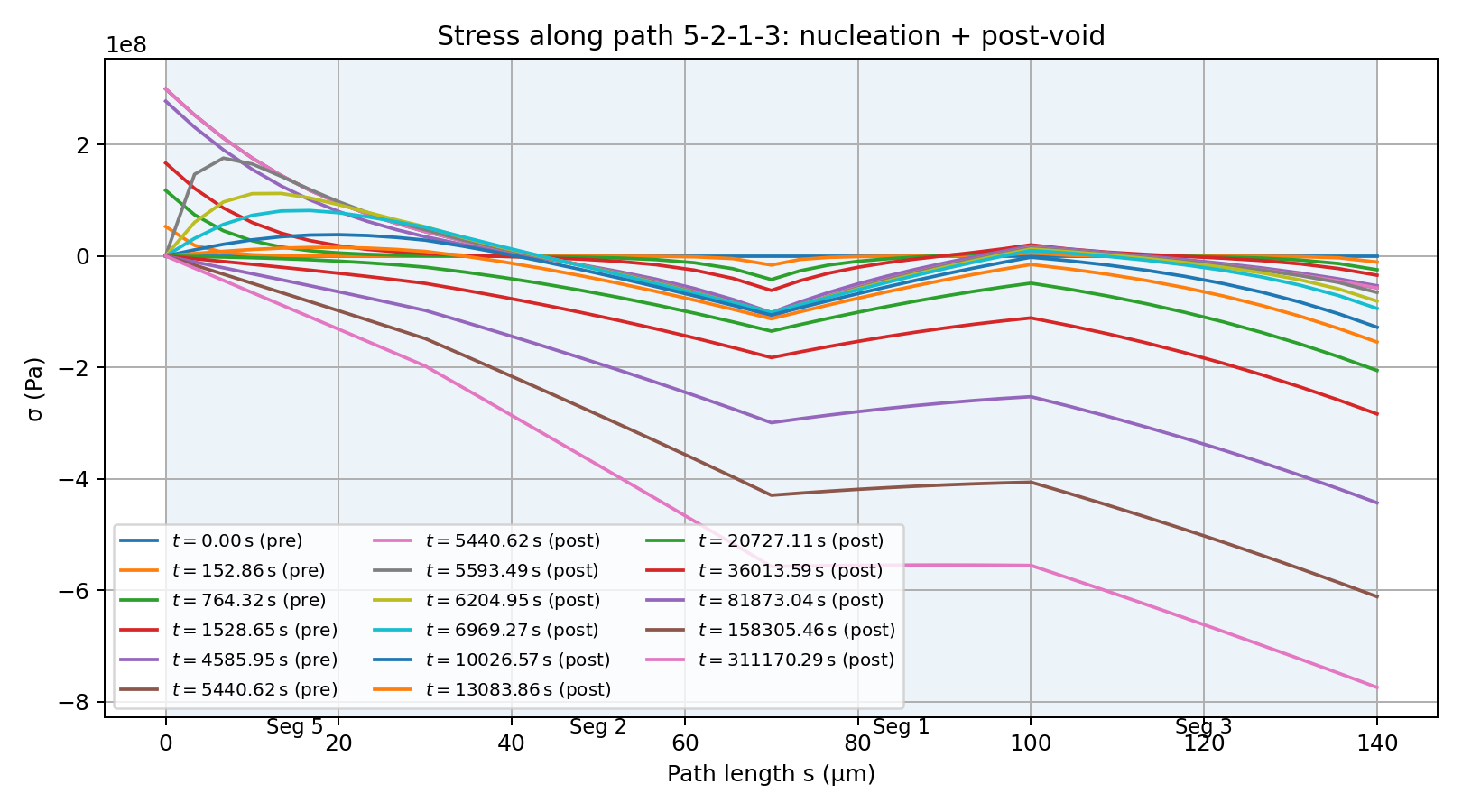}
  \vspace{-5pt}
  \caption{\small Stress evolution in four segments of a five-segment wire, including the nucleation and post-void phases} 
  \label{fig:stress_evolution_wire0_5seg}
  \vspace{-10pt}
\end{figure}

The additional BCs that describe the post-voiding phase are outlined in Eq.~\eqref{eq:korhonen_equation_single_wire_postvoid}~\cite{sukharev2016postvoiding}, which assumes that the void forms at the cathode node $x=x_{nuc}$. 
\begin{equation}
\small
    \begin{aligned}
        % &PDE:  \frac{\partial \sigma(x,t)}{\partial t}=\frac{\partial
        %   }{\partial x}\left[\kappa(\frac{\partial \sigma(x,t)}{\partial x} + G)\right],\ t> 0  \\ 
        &BC_{robin}: \frac{\partial \sigma(x_{nuc},t)}{\partial x} \big |_{x=x_{nuc}} = \frac{\sigma(x_{nuc},t)}{\delta},\ t > 0 \\
        &BC_{neu}: \frac{\partial \sigma(0,t)}{\partial x} \big |_{x=x_b} = -G ,\ t > 0 \\ 
        &IC: \sigma (x,0) =  \sigma_{nuc}(x,t_{nuc}),\ t=0 
   \label{eq:korhonen_equation_single_wire_postvoid}
    \end{aligned}
\end{equation}
Here, $\delta$ is the effective void thickness, which is typically very small. As a result, the Robin BC $\frac{\partial \sigma(x_{nuc},t)}{\partial x} \big |_{x=x_{nuc}} = \frac{\sigma(x_{nuc},t)}{\delta}$ rapidly drives the stress at $x_{nuc}$ toward zero, and the atomic flux at that location also quickly vanishes. The stresses in all wire segments eventually become negative at steady state, corresponding to compressive stress. Fig.~\ref{fig:stress_evolution_wire0_5seg} shows the stress evolution in four segments of a five-segment wire. The void nucleates at the terminal of segment 5, with a critical stress of $3\times10^8$ Pa. Almost immediately after nucleation, the terminal stress drops to zero. 

\subsection{Finite-Difference Method for Solving the EM PDE}

Using the finite-difference method, Eq.~\eqref{eq:korhonen_multisegment_nucleation}
is discretized by splitting each segment into uniformly spaced points with spacing $\Delta x$. Approximating the spatial derivative by its finite-difference form, for each discretized point $i$ we obtain:
\begin{equation}
\small
\frac{\partial \sigma}{\partial t} = \kappa \frac{\partial^2 \sigma}{\partial x^2}  = \kappa\frac{\sigma_{i+1} - 2\sigma_i
      + \sigma_{i-1}}{\Delta x^2}.
\label{eq:korhonen_simple}
\end{equation}
for all internal nodes. 

% Applying finite differences to an $k$-segment interconnect wire will lead to  in $n$ discretized points, which may correspond to
% various topological positions: left and right boundaries, middle
% segment points, intermediate junctions, and via junctions.
% Considering the boundary conditions in 
% Eq.~\eqref{eq:korhonen_multisegment_nucleation} leads to the following
% linear ordinary differential equation (ODE) system:
% \begin{equation}
% C \dot{\boldsymbol{\sigma}}(t)
%   = G \boldsymbol{\sigma}(t) + B \boldsymbol{j}(t),
% \label{eq:lti_em}
% \end{equation}
% which describes the EM stress evolution for each interconnect segment and can be expressed as a
% Linear Time-Invariant (LTI) system.
Applying finite-difference discretization to a $k$-segment interconnect wire
yields $n$ discrete nodes, which represent different topological locations along the wire tree, such as natural boundaries (including via connections), inter-segment points, and inter-segment junctions. Incorporating the boundary
conditions specified in Eq.~\eqref{eq:korhonen_multisegment_nucleation}
results in the following system of linear ordinary differential equations:
\begin{equation}
\small
C \dot{\boldsymbol{\sigma}}(t)
  = A \boldsymbol{\sigma}(t) + B \boldsymbol{j}(t),
\label{eq:lti_em}
\end{equation}
which governs the evolution of electromigration-induced stress within each
interconnect segment and can be formulated as a linear time-invariant (LTI)
system. Once the stress reaches the critical stress, a void will be nucleated at the cathode node of the wire segment. The void volume $V_v(t)$ in a multi-segment interconnect is governed by atom
conservation, relating hydrostatic stress to the amount of depleted material~\cite{Korhonen:1993bb}:
\begin{equation}
\small
V_v(t) \;=\; - \int_{\Omega_L} \frac{\sigma(t)}{B}\, dV,
\label{eq:emspice_Vv}
\end{equation}
where $\Omega_L$ is the remaining wire domain, $\sigma(t)$ is the hydrostatic
stress field, and $B$ is the (effective) bulk modulus of the metal. Once the void exceeds the critical size, such as the wire width, current is partially
diverted through the barrier, and the wire resistance increases, which can be approximated as
\begin{equation}
\small
\Delta R(t) \;=\; \frac{V_v(t) - V_{\mathrm{crit}}}{W H}
\left[
  \frac{\rho_{\mathrm{Ta}}}{h_{\mathrm{Ta}}(2H+W)}
  \;-\;
  \frac{\rho_{\mathrm{Cu}}}{H W}
\right],
\, \, t_{\mathrm{inc}} < t,
\label{eq:emspice_dR}
\end{equation}
where $W$ and $H$ are the copper line width and thickness, $\rho_{\mathrm{Ta}}$
and $\rho_{\mathrm{Cu}}$ are the resistivities of barrier (e.g., Ta/TaN) and Cu,
and $h_{\mathrm{Ta}}$ is the barrier thickness. Here $V_{\mathrm{crit}}$ denotes
the critical void volume (or an equivalent critical cross-section length in 1D).

\subsection{Overview of the Rational Krylov Subspace Method}

For the descriptor-form stress system
\begin{equation}
\small
C\dot{\boldsymbol{\sigma}}(t)=A\boldsymbol{\sigma}(t)+B\boldsymbol{j}(t),
\label{eq:lti_em_overview}
\end{equation}
the classical Krylov subspace method constructs an orthonormal basis from the
propagation operator $G=A^{-1}C$ and the source block
$\mathcal{B}=[A^{-1}C\boldsymbol{\sigma}_0,\;A^{-1}B]$:
\begin{equation}
\small
\mathcal{K}_q(G,\mathcal{B})
=\mathrm{range}\!\left(
\left[\,
\mathcal{B},\,
G\mathcal{B},\,
\ldots,\,
G^{q-1}\mathcal{B}
\right]
\right).
\end{equation}
This standard Krylov basis emphasizes the behavior around the expansion point
$s = 0$ or, equivalently, $t = \infty$. As a result, its accuracy deteriorates
when the important dynamics occur at earlier times or other frequency ranges.

To overcome this limitation, the \emph{rational} Krylov subspace introduces a
shift or expansion point $s_0 \in \mathbb{C}$. Defining
$K_{s_0}=s_0 C-A$ and $R_{s_0}=K_{s_0}^{-1}$, the shifted block rational
Krylov subspace is
\begin{equation}
\small
\begin{aligned}
\mathcal{K}_q\!\big(R_{s_0}C,\mathcal{B}_{s_0}\big)
= {} & \mathrm{range}\!\left(
\left[\,
\mathcal{B}_{s_0},\,
(R_{s_0}C)\mathcal{B}_{s_0},\,
\ldots,\right.\right.\\
& \left.\left.
(R_{s_0}C)^{q-1}\mathcal{B}_{s_0}
\right]
\right),
\end{aligned}
\end{equation}
where
\begin{equation}
\small
\mathcal{B}_{s_0}
=
\left[\,
R_{s_0}C\boldsymbol{\sigma}_0,\;
R_{s_0}B
\right].
\end{equation}

Let $V_q = [\boldsymbol{v}_1, \ldots, \boldsymbol{v}_q] \in \mathbb{R}^{n\times q}$
be the orthonormal basis of this subspace. The rational Arnoldi
decomposition is given by:
\begin{equation}
\small
(s_0 C-A)^{-1} C V_q
= V_q H_q + h_{q+1,q}\boldsymbol{v}_{q+1} e_q^{\top},
\label{eq:arnoldi_decomposition}
\end{equation}
where $H_q \in \mathbb{R}^{q\times q}$ is an upper Hessenberg matrix and
$e_q^{\top} = [0,\,\ldots,\,0,\,1]$.

By shifting the system matrix with $(s_0 C-A)$, the resulting basis better
aligns with the spectral region corresponding to the time window or frequency
range of interest, such as the nucleation time in EM stress evolution. The
rational Krylov process therefore enables accurate reduced-order models with
significantly smaller dimensions, because the choice of $s_0$ focuses the
projection space on the most relevant poles of the system. The final reduced
matrices are obtained by congruence transformation:
\begin{equation}
\small
\hat{A} = V_q^{T} A V_q, \qquad
\hat{B} = V_q^{T} B, \qquad
\hat{C} = V_q^{T} C V_q.
\end{equation}
This framework generalizes the standard Krylov approach and serves as the
foundation for our proposed extended rational Krylov reduction method for fast
EM analysis.

\section{EM Stress Analysis Based on the Extended Rational Krylov Subspace Method}
\label{sec:extrakrylov_method}

\subsection{Proposed Extended Rational Krylov EM Analysis}

After spatial discretization of the electromigration (EM) stress equation,
the resulting linear time-invariant (LTI) system can be written as
\begin{equation}
\small
C \dot{\boldsymbol{\sigma}}(t) = A \boldsymbol{\sigma}(t) + B \boldsymbol{j}(t),
\qquad
\boldsymbol{\sigma}(0) = \boldsymbol{\sigma}_0,
\label{eq:lti_identity_rev}
\end{equation}
where $\boldsymbol{\sigma}(t)$ denotes the nodal stress vector and
$\boldsymbol{j}(t)$ is the applied current-density input.

\subsubsection{Piecewise-Linear Current Model}

We assume the current density is piecewise linear in time with breakpoints
$0=t_0<t_1<\cdots<t_N$ and values $\boldsymbol{j}_k=\boldsymbol{j}(t_k)$.
On each interval $[t_{k-1},t_k)$,
\[
\boldsymbol{j}(t)=\boldsymbol{j}_{k-1}+\boldsymbol{r}_k (t-t_{k-1}),
\qquad
\boldsymbol{r}_k=\frac{\boldsymbol{j}_k-\boldsymbol{j}_{k-1}}{t_k-t_{k-1}}.
\]
A global ramp representation using shifted ramps is
\begin{equation}
\small
\begin{aligned}
\boldsymbol{j}(t)
= {} & \boldsymbol{j}_0
+\sum_{k=1}^{N}\boldsymbol{r}_k(t-t_{k-1})H(t-t_{k-1}) \\
& -\sum_{k=1}^{N}\boldsymbol{r}_k(t-t_k)H(t-t_k),
\end{aligned}
\label{eq:pwl_j}
\end{equation}
where $H(\cdot)$ is the unit step. Using
$\mathcal{L}\{(t-a)H(t-a)\}=e^{-as}/s^2$, the Laplace transform is
\begin{equation}
\small
\begin{aligned}
\boldsymbol{J}(s)
= {} & \frac{\boldsymbol{j}_0}{s}
+\sum_{k=1}^{N}\boldsymbol{r}_k
\left(
\frac{e^{-s t_{k-1}}}{s^2}
-\frac{e^{-s t_k}}{s^2}
\right),
\end{aligned}
\label{eq:J_pwl}
\end{equation}

\subsubsection{Shifted Expansion Using $\widetilde{\Sigma}(s)=s^2\Sigma(s)$}

Applying the Laplace transform to~\eqref{eq:lti_identity_rev} gives
\begin{equation}
\small
(sC-A)\Sigma(s)=C\boldsymbol{\sigma}_0 + B \boldsymbol{J}(s).
\label{eq:laplace_rev}
\end{equation}
Because the piecewise-linear input contains $1/s^2$ factors, we define
\begin{equation}
\small
\widetilde{\Sigma}(s):=s^2\Sigma(s),
\label{eq:Sigma_tilde_def}
\end{equation}
which removes the singular behavior at $s=0$ in the input contribution.
Multiplying~\eqref{eq:laplace_rev} by $s^2$ and substituting
$\Sigma(s)=\widetilde{\Sigma}(s)/s^2$ yields
\begin{equation}
\small
\begin{aligned}
(sC-A)\widetilde{\Sigma}(s)
= {} & s^{2} C \boldsymbol{\sigma}_0 + s^{2} B \boldsymbol{J}(s).
\end{aligned}
\label{eq:laplace_tilde_sigma}
\end{equation}

We now expand around the shifted frequency $s=s_0$. Let
\[
s=s_0+\Delta,\qquad
K_{s_0}:=(s_0 C-A),\qquad
R_{s_0}:=K_{s_0}^{-1}.
\]
Substituting $s=s_0+\Delta$ into~\eqref{eq:laplace_tilde_sigma} gives
\begin{equation}
\small
\begin{aligned}
\big(K_{s_0}+\Delta C\big)\,\widetilde{\Sigma}(s_0+\Delta)
= {} & (s_0+\Delta)^2 C \boldsymbol{\sigma}_0 \\
& + (s_0+\Delta)^2 B \boldsymbol{J}(s_0+\Delta).
\end{aligned}
\label{eq:shifted_equation}
\end{equation}

We expand $\widetilde{\Sigma}(s_0+\Delta)$ and the transformed source term into
power series:
\begin{equation}
\small
\begin{aligned}
\widetilde{\Sigma}(s_0+\Delta)
&= \sum_{k=0}^{\infty}\widetilde{m}_k\,\Delta^k,\\
(s_0+\Delta)^2\boldsymbol{J}(s_0+\Delta)
&= \sum_{k=0}^{\infty} w_k\,\Delta^k.
\end{aligned}
\label{eq:series_tilde_sigma}
\end{equation}
For the piecewise-linear input~\eqref{eq:J_pwl}, multiplying by $s^2$ cancels
the $1/s^2$ factors and yields
\[
s^2\boldsymbol{J}(s)=s\boldsymbol{j}_0+\sum_{i=1}^{N}\boldsymbol{r}_i
\big(e^{-s t_{i-1}}-e^{-s t_i}\big).
\]
Using
\[
e^{-(s_0+\Delta)t}=e^{-s_0 t}\sum_{k=0}^{\infty}\frac{(-t)^k}{k!}\Delta^k
\]
gives the source moments
\begin{equation}
\small
\begin{aligned}
w_k
= {} & \delta_{k0}\,s_0\boldsymbol{j}_0
+\delta_{k1}\,\boldsymbol{j}_0 \\
& +\sum_{i=1}^{N}\boldsymbol{r}_i
\left(
e^{-s_0 t_{i-1}}\frac{(-t_{i-1})^k}{k!}
-
e^{-s_0 t_i}\frac{(-t_i)^k}{k!}
\right),
\end{aligned}
\label{eq:wk}
\end{equation}
where $\delta_{kj}$ is the Kronecker delta. Substituting
~\eqref{eq:series_tilde_sigma} into~\eqref{eq:shifted_equation} and matching
equal powers of $\Delta$ yields the corrected shifted moments:

\paragraph{Zeroth-order ($\Delta^0$):}
\begin{equation}
\small
\begin{aligned}
K_{s_0}\widetilde{m}_0
= {} & s_0^2 C\boldsymbol{\sigma}_0 + B w_0,\\
\widetilde{m}_0
= {} & R_{s_0}\!\left(s_0^2 C\boldsymbol{\sigma}_0 + B w_0\right).
\end{aligned}
\label{eq:m0}
\end{equation}

\paragraph{Higher orders ($\Delta^k$, $k\ge 1$):}
\begin{equation}
\small
K_{s_0}\widetilde{m}_k + C\widetilde{m}_{k-1}
= C\alpha_k\boldsymbol{\sigma}_0 + B w_k,
\label{eq:mk_recursion}
\end{equation}
\begin{equation}
\small
\begin{aligned}
\widetilde{m}_k
= {} & R_{s_0}\!\left(
C\alpha_k\boldsymbol{\sigma}_0 + B w_k - C\widetilde{m}_{k-1}
\right),\\
& k\ge 1,
\end{aligned}
\label{eq:mk_explicit}
\end{equation}
where $\alpha_0=s_0^2$, $\alpha_1=2s_0$, $\alpha_2=1$, and $\alpha_k=0$ for
$k\ge 3$.

Equations~\eqref{eq:m0}--\eqref{eq:mk_explicit} show that the previous moment
propagates through the operator $R_{s_0}C$, while the source moments enter
through $R_{s_0}B$. Accordingly, the correct approximation space is a block
rational Krylov space rather than a single-vector Krylov chain. Define the
starting block
\begin{equation}
\small
\mathcal{B}_{s_0}:=
\left[\,
R_{s_0} C\boldsymbol{\sigma}_0,\;
R_{s_0} B
\right],
\label{eq:b_sigma}
\end{equation}
and the shifted block rational Krylov subspace
\begin{equation}
\small
\begin{aligned}
\mathcal{K}_q(R_{s_0}C,\mathcal{B}_{s_0})
= {} & \mathrm{range}\!\left(
\left[\,
\mathcal{B}_{s_0},\;
(R_{s_0}C)\mathcal{B}_{s_0},\;
\right.\right. \\
& \left.\left.
(R_{s_0}C)^2\mathcal{B}_{s_0},\;
\ldots,\;
(R_{s_0}C)^{q-1}\mathcal{B}_{s_0}
\right]
\right).
\end{aligned}
\label{eq:rational_krylov}
\end{equation}
All shifted moments lie in this subspace, which yields good transient accuracy
around the time scale implied by $s_0$. If the initial stress is zero, the
starting block reduces to $R_{s_0}B$. The orthonormal basis $V_q$ is
constructed via a block Arnoldi process as shown in
Algorithm~\ref{alg:ExtRationalArnoldi}.

\subsubsection{Reduced-Order Model and Time-Domain Simulation}

Using the projection matrix $V_q$ formed by the orthonormal basis vectors, the
reduced matrices are
\begin{align}
\hat{A} &= V_q^{T} A V_q, &
\hat{B} &= V_q^{T} B, &
\hat{C} &= V_q^{T} C V_q.
\end{align}
The reduced descriptor system is
\begin{equation}
\small
\begin{aligned}
\hat{C}\dot{\hat{\boldsymbol{\sigma}}}(t)
= {} & \hat{A}\hat{\boldsymbol{\sigma}}(t)+\hat{B}\boldsymbol{j}(t),\\
\hat{\boldsymbol{\sigma}}(0)
= {} & V_q^{T}\boldsymbol{\sigma}_0.
\end{aligned}
\label{eq:reduced_rom}
\end{equation}
Backward-Euler integration is applied to the reduced system. In the shifted
moment domain, the reduced source moments enter through the same sequence
$w_k$, i.e.,
\begin{equation}
\small
\begin{aligned}
\hat{K}_{s_0}\hat{\widetilde{m}}_0
= {} & s_0^2 \hat{C}\hat{\boldsymbol{\sigma}}_0 + \hat{B} w_0,\\
\hat{K}_{s_0}\hat{\widetilde{m}}_k + \hat{C}\hat{\widetilde{m}}_{k-1}
= {} & \hat{C}\alpha_k\hat{\boldsymbol{\sigma}}_0 + \hat{B} w_k,
\end{aligned}
\label{eq:reduced_moment_recursion}
\end{equation}
where $\hat{K}_{s_0}=s_0\hat{C}-\hat{A}$ and
$\hat{\boldsymbol{\sigma}}_0=V_q^{T}\boldsymbol{\sigma}_0$. The full-space
stress field is recovered by
\begin{equation}
\small
\boldsymbol{\sigma}(t)\approx V_q \hat{\boldsymbol{\sigma}}(t).
\end{equation}

\subsubsection{Constant-Current Special Case}
If the input current is constant, i.e., $\boldsymbol{j}(t) = \boldsymbol{j}_0$ for all $t$, then $\boldsymbol{J}(s)=\boldsymbol{j}_0/s$ and
$s^2\boldsymbol{J}(s)=s\boldsymbol{j}_0$. Expanding around $s=s_0+\Delta$ gives
$w_0=s_0\boldsymbol{j}_0$, $w_1=\boldsymbol{j}_0$, and $w_k=0$ for $k\ge 2$. The moment recursion in Eq.~\eqref{eq:reduced_moment_recursion} then simplifies to
\begin{equation}
\small
\begin{aligned}
\hat{K}_{s_0}\hat{\widetilde{m}}_0 &= s_0^2 \hat{C}\hat{\boldsymbol{\sigma}}_0 + \hat{B} w_0, \\
\hat{K}_{s_0}\hat{\widetilde{m}}_1 + \hat{C}\hat{\widetilde{m}}_0 &= \hat{C}\alpha_1\hat{\boldsymbol{\sigma}}_0 + \hat{B} w_1, \\
\hat{K}_{s_0}\hat{\widetilde{m}}_k + \hat{C}\hat{\widetilde{m}}_{k-1} &= \hat{C}\alpha_k\hat{\boldsymbol{\sigma}}_0, \quad k \ge 2.
\end{aligned}
\end{equation}
Thus, only the first two source moments are nonzero, and all higher-order recursions depend only on the initial stress and the system matrices.

In this paper, we assume that the current is constant during both nucleation and post-void phases, which is a common scenario in EM analysis. However, the proposed {\it ExtRaKrylovEM} method is general and can handle arbitrary time-varying piecewise-linear current profiles.

\subsubsection{Remark on Singularity Mitigation}

Because we solve systems with the shifted operator $(s_0 C-A)$ rather than
inverting $A$, the method avoids direct singularity/near-singularity issues of
$A$ under Neumann-type constraints as long as $s_0$ is chosen outside the
generalized spectrum of the matrix pair $(A,C)$, which mitigates the
singularity concern reported in~\cite{Chatterjee:2016ICCAD,CookSun:TVLSI'18}.

\begin{algorithm}[t]
\caption{Source-Moment-Aware Block Rational Krylov Reduction}
\label{alg:ExtRationalArnoldi}
\small
\begin{algorithmic}[1]
\Require System matrices $A$, $B$, $C$, initial stress $\boldsymbol{\sigma}_0$,
         PWL current data $\{\boldsymbol{j}_k,t_k\}_{k=0}^{N}$,
         expansion point $s_0$, and reduction order $q$
\Ensure Reduced matrices $A_h$, $B_h$, $C_h$, projection matrix $V$,
        source moments $\{w_k\}_{k=0}^{q-1}$, and
        reduced shifted moments $\{\hat{\widetilde{m}}_k\}_{k=0}^{q-1}$
\vspace{0.5em}
\For{$k = 1$ to $N$}
    \State $\boldsymbol{r}_k \leftarrow
    (\boldsymbol{j}_k-\boldsymbol{j}_{k-1})/(t_k-t_{k-1})$
\EndFor
\For{$m = 0$ to $q-1$}
    \State Compute the source moment $w_m$ using \eqref{eq:wk}
\EndFor
\State Form $K_{s_0} \leftarrow s_0 C - A$ and compute its LU factors
\State Define $\mathcal{M}(X) \leftarrow K_{s_0}^{-1} C X$
\State Form $F_0 \leftarrow
\left[K_{s_0}^{-1} C\boldsymbol{\sigma}_0,\; K_{s_0}^{-1} B\right]$
\State Compute $F_0 = V_1 R_0$ and initialize $V \leftarrow V_1$
\vspace{0.5em}
\For{$j = 1$ to $q-1$}
    \State $W \leftarrow \mathcal{M}(V_j)$
    \For{$i = 1$ to $j$}
        \State $H_{ij} \leftarrow V_i^{T} W$
        \State $W \leftarrow W - V_i H_{ij}$
    \EndFor
    \State Compute the thin QR factorization $W = V_{j+1} H_{j+1,j}$
    \If{$\|H_{j+1,j}\|_F < \varepsilon$} \textbf{break}
    \EndIf
    \State Append $V_{j+1}$ to $V$
\EndFor
\vspace{0.5em}
\State Form reduced matrices:
\[
A_h = V^T A V,\quad
B_h = V^T B,\quad
C_h = V^T C V
\]
\State $\hat{K}_{s_0} \leftarrow s_0 C_h - A_h$
\State $\hat{\boldsymbol{\sigma}}_0 \leftarrow V^T\boldsymbol{\sigma}_0$
\State $\hat{\widetilde{m}}_0 \leftarrow
\hat{K}_{s_0}^{-1}(s_0^2 C_h\hat{\boldsymbol{\sigma}}_0 + B_h w_0)$
\For{$m = 1$ to $q-1$}
    \State $\hat{\widetilde{m}}_m \leftarrow \hat{K}_{s_0}^{-1}(
    C_h\alpha_m\hat{\boldsymbol{\sigma}}_0 + B_h w_m
    - C_h\hat{\widetilde{m}}_{m-1})$
\EndFor
\State \Return $(A_h, B_h, C_h, V, \{w_k\}_{k=0}^{q-1},
\{\hat{\widetilde{m}}_k\}_{k=0}^{q-1})$
\end{algorithmic}
\end{algorithm}

% \begin{algorithm}[t]
% \caption{Extended Rational Arnoldi Process for FastEM Reduction}
% \label{alg:ExtRationalArnoldi}
% \begin{algorithmic}[1]
% \Require System matrices $A$, $B$, $C$, input vector $u$, 
%          rational shift $\sigma$, optional initial shift $\sigma_0$, 
%          and reduction order $m$
% \Ensure Reduced matrices $A_h$, $B_h$, $C_h$, projection matrix $V$, and upper Hessenberg matrix $H$
% \vspace{0.5em}
% \State Factorize the shifted matrix: $(A - \sigma I) = LU$
% \State Define the inverse operator $\mathcal{A}^{-1}v = (A - \sigma I)^{-1}v$
% \State Compute $r_1 = -\mathcal{A}^{-1}(Bu)$
% \State Normalize $v_1 = r_1 / \|r_1\|$, set $V = [v_1]$
% \vspace{0.5em}
% \If{$\sigma_0 \neq 0$}
%     \State $r_2 = \mathcal{A}^{-1}(C(v_1 - \sigma_0))$
% \Else
%     \State $r_2 = \mathcal{A}^{-1}(Cv_1)$
% \EndIf
% \State Orthogonalize: $r_2 = r_2 - (v_1^T r_2)v_1$, normalize $v_2 = r_2 / \|r_2\|$
% \State Append $v_2$ to $V$
% \vspace{0.5em}
% \For{$j = 3$ to $m$}
%     \State $w = \mathcal{A}^{-1}(v_{j-1})$
%     \For{$i = 1$ to $j-1$}
%         \State $h_{ij} = v_i^T w$
%         \State $w = w - h_{ij} v_i$
%     \EndFor
%     \State $\beta = \|w\|$
%     \If{$\beta < \varepsilon$} \textbf{break}
%     \EndIf
%     \State $v_j = w / \beta$
%     \State Append $v_j$ to $V$
% \EndFor
% \vspace{0.5em}
% \State Form reduced matrices: 
% \[
% A_h = V^T A V,\quad 
% B_h = V^T B,\quad 
% C_h = V^T C V
% \]
% \State \Return $(A_h, B_h, C_h, V, H)$
% \end{algorithmic}
% \end{algorithm}

\subsection{Selection of reduction order and shift time}

One critical aspect of the rational Krylov subspace method is selecting the
expansion point $s_0$. In this work, instead of using $s_0$ directly, we use
shift time $t_{s} = 1/s_0$, which can be viewed as the time constant of the
system behavior of greatest interest. Two questions naturally arise: (i) how
to obtain a good initial estimate for the shift time, and (ii) how to
optimally tune it along with the reduction order. The following two subsections
address each in turn.

\subsection{Initial Estimation of Shift Time for the Nucleation and Post-Void Phases}

To estimate the initial shift time for the nucleation phase, we use the analytical solution of the one-dimensional diffusion equation as a reference. The 1D diffusion equation is
\begin{equation}
\frac{\partial \sigma(x,t)}{\partial t}
= \kappa \,\frac{\partial^2 \sigma(x,t)}{\partial x^2}
\end{equation}
where $\sigma(x,t)$ is the hydrostatic stress and $\kappa$ is the effective stress diffusivity. For a given wire segment of length $L$ under atomic blocking conditions at both ends, the system has the following eigenvalues:
\begin{equation}
\lambda_n = \left(\frac{n\pi}{L}\right)^2,\quad n=1,2,\dots
\end{equation}
By using the slowest decaying mode where $n=1$,
$
\lambda_1 = \left(\frac{\pi}{L}\right)^2
$.
So the slowest exponential decay is:
$
\exp\!\left(-\kappa \frac{\pi^2}{L^2} t\right)
$
Therefore, the shift time can be estimated as
\begin{equation}
\tau = \frac{L^2}{\pi^2 \kappa}
\label{eq:shift_time_estimation}
\end{equation}
We estimate the shift time using Eq.~\eqref{eq:shift_time_estimation}: for the nucleation phase, we set the base time $\tau_{\mathrm{nuc}}$ by substituting the average wire-segment length $L_{\text{avg}}$, and for the post-void phase, we set the base time $\tau_{\mathrm{post}}$ by substituting the maximum path length $L_{\text{max}}$ of the interconnect tree. These initial estimates are then refined in the next subsection to select the best shift times for both phases.
This choice is motivated by the distinct physics governing each phase. Nucleation time is predominantly influenced by local stress gradients within individual wire segments under electron-wind forcing, making it sensitive to the characteristic segment length. In contrast, post-void resistance evolution is governed by void propagation along the critical path of the interconnect tree, rendering it sensitive to the maximum path length over which atom depletion occurs. This dual-scale approach ensures that the shift times reflect the dominant time constants in each regime of EM degradation. For the nucleation-phase shift, we can also use the analytical nucleation-time formula for a single wire segment~\cite{Huang:TCAD'15}. Since these shift times are only initial estimates, either choice is acceptable.

\subsection{Coordinate Descent Algorithm for Order and Shift Time Selection}

In this subsection, we present a coordinate descent strategy with multi-scale line search to tune the Krylov order and shift parameters. The optimization focuses on the reduction order and shift times for both nucleation and post-void phases in the rational Krylov EM analysis. The goal is to identify the optimal configuration that minimizes the combined percentage error in nucleation time and resistance change within the specified parameter search ranges, while maintaining the reduction order as low as possible to ensure computational efficiency.

\begin{figure}[h]
  \vspace{-5pt}
  \centering
  \includegraphics[width=0.85\columnwidth]{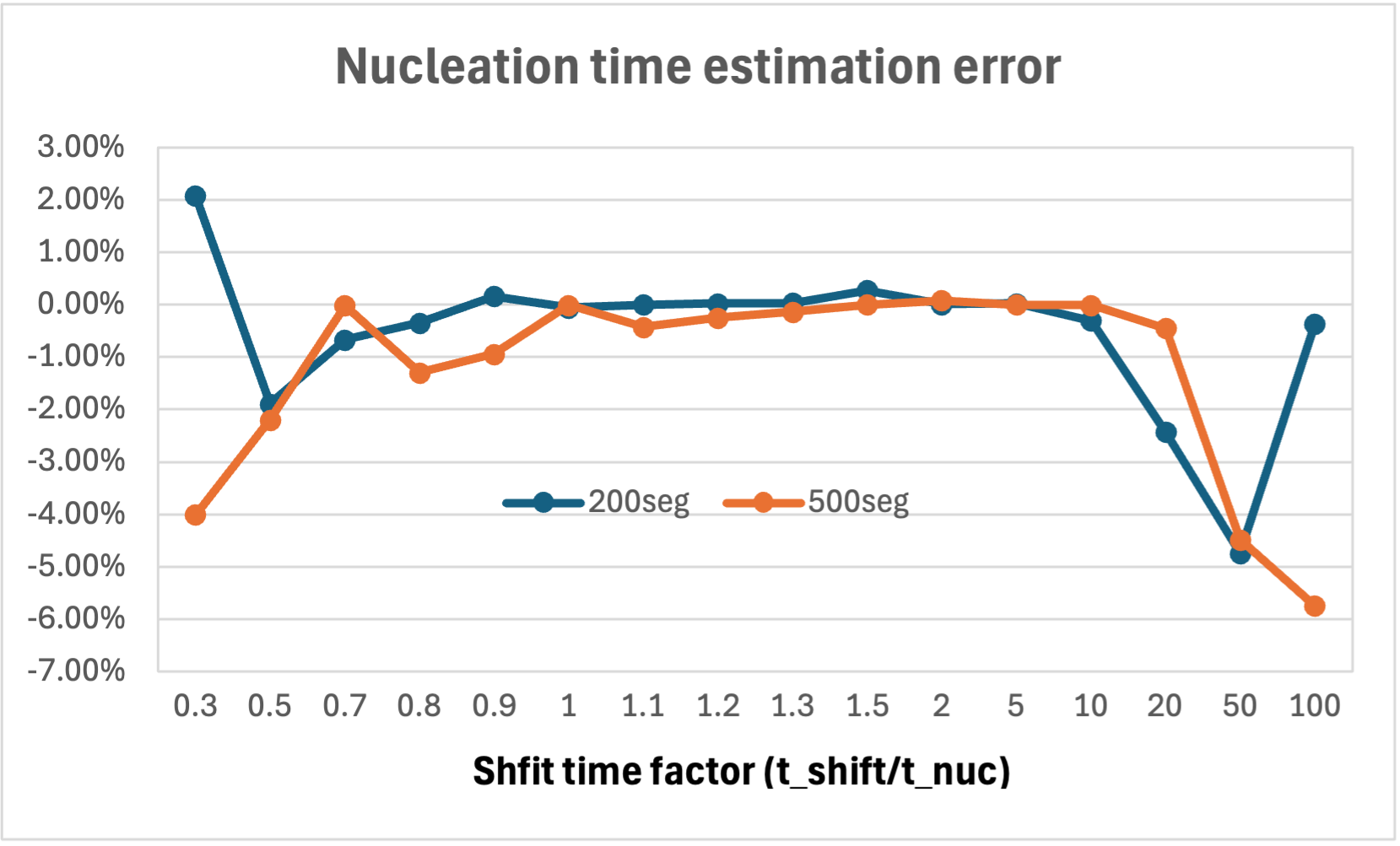}
  \vspace{-5pt}
  \caption{\small Calculated nucleation time versus the shift time used (critical stress $= 10^8$ Pa)} 
  \label{fig:nuc_time_vs_shift_time}
  \vspace{-5pt}
\end{figure}

In the coordinate-descent algorithm, we use the nucleation time $t_{nuc}$ and the resistance change $\Delta R$ of the nucleated wire segment as performance metrics and compute their percentage errors with respect to the reference results. We vary the parameters one at a time to avoid evaluating objective gradients directly. This strategy is well suited to our optimization problem because the number of parameters is small and the search space is limited. 
 
Furthermore, instead of using the shift times directly, we introduce shift-time scaling factors $\eta_{\mathrm{nuc}}$ and $\eta_{\mathrm{post}}$ for the nucleation and post-void phases, respectively. For instance, $\eta_{\mathrm{nuc}} = 1$ means that the estimated shift time for the nucleation phase is $1 \times \tau_{\mathrm{nuc}}$, while other values scale it accordingly. Similarly, $\eta_{\mathrm{post}}$ scales the shift time for the post-void phase.
 
Next, we determine the search ranges for these parameters. Fig.~\ref{fig:nuc_time_vs_shift_time} illustrates the nucleation-time estimation errors obtained by the ExtRaKrylov method across different shift-time factors $\eta_{\mathrm{nuc}}$ ranging from 0.3 to 100. As shown, when $\eta_{\mathrm{nuc}}$ lies between 0.7 and 10, the results exhibit high accuracy (less than 0.01\% error on average). In practice, setting $\eta_{\mathrm{nuc}} = 1$ typically achieves less than 1\% error with order 4 or 5. Similar behavior is observed for post-void resistance-change estimation. Nevertheless, we develop a coordinate-descent method to systematically identify the optimal combination of order and shift time for both the nucleation and post-void phases.

Based on these observations, we set the candidate order set or range to $\mathcal{M} = \{3,4,5,6\}$, since orders within this range typically provide sufficient accuracy. For the shift time scaling factors $\eta_{\mathrm{nuc}}$ and $\eta_{\mathrm{post}}$, we set their search ranges to $[0.1, 20]$, as shift times outside this range yield poor accuracy as shown in Fig.~\ref{fig:nuc_time_vs_shift_time}.

Algorithm~\ref{alg:deep_descent_krylov} summarizes the proposed coordinate-descent approach for tuning rational Krylov parameters in EM analysis. The algorithm seeks to minimize a combined objective function $J$ that captures both nucleation-time and resistance-change errors. The same algorithm is also applied to the {\it EiRaKrylovEM} solver for the nucleation and post-void phases. We define the percentage error (PE) as
\begin{equation}
\mathrm{PE}(x,y) \triangleq 100\frac{|x-y|}{|x|+\epsilon},
\end{equation}
where $\epsilon$ is a small regularization constant to avoid division by zero. The combined objective function is then given by
\begin{equation}
\begin{aligned}
J(q,\eta_{\mathrm{nuc}},\eta_{\mathrm{post}}) \triangleq 
&\mathrm{PE}\!\left(t_{\mathrm{nuc}}^{\mathrm{ref}},\, t_{\mathrm{nuc}}(q,\eta_{\mathrm{nuc}})\right) \\
&+\mathrm{PE}\!\left(\Delta R^{\mathrm{ref}},\, \Delta R(q,\eta_{\mathrm{post}})\right),
\end{aligned}
\end{equation}
where $t_{\mathrm{nuc}}^{\mathrm{ref}}$ and $\Delta R^{\mathrm{ref}}$ denote reference values from high-fidelity simulations, while $t_{\mathrm{nuc}}(q,\eta_{\mathrm{nuc}})$ and $\Delta R(q,\eta_{\mathrm{post}})$ represent the reduced-order predictions for nucleation time and resistance change, respectively.

The main routine performs coordinate descent to optimize the reduction order $q$ and the shift-time scaling factors $\eta_{\mathrm{nuc}}$ and $\eta_{\mathrm{post}}$. The algorithm iteratively refines these parameters by alternating discrete searches over the candidate order set $\mathcal{M}$ with step-based searches for the shift factors within their bounds $[\eta_{\min},\eta_{\max}]$. Iterations continue until the improvement in $J$ falls below a threshold $\tau_{\mathrm{stop}}$ or the maximum number of iterations is reached. The optimized parameters $(q^\star,\eta_{\mathrm{nuc}}^\star,\eta_{\mathrm{post}}^\star)$ and the corresponding minimal objective $J^\star$ are then used in the rational Krylov EM solvers.

\begin{algorithm}[t]
\caption{Coordinate Descent Method For Shift Time and Order Selection}
\label{alg:deep_descent_krylov}
\begin{algorithmic}[1]
\Require candidate order set $\mathcal{M}$, shift-factor bounds $[\eta_{\min},\eta_{\max}]$
\Ensure Best parameters $(q^\star,\eta_{\mathrm{nuc}}^\star,\eta_{\mathrm{post}}^\star)$ and best $J^\star$

\vspace{2pt}
\Statex \textbf{Initialization}
\State Choose initial $(q,\eta_{\mathrm{nuc}},\eta_{\mathrm{post}})$, $J^\star \leftarrow J(q,\eta_{\mathrm{nuc}},\eta_{\mathrm{post}})$
\State Set step-size list $\mathcal{S}$ (coarse-to-fine)

\vspace{2pt}
\Statex \textbf{Coordinate Descent Loop}
\For{$iter = 1$ to $I_{\max}$}
    \State $J_{\mathrm{prev}} \leftarrow J^\star$

    \Statex \hspace{0.5em}\emph{(1) Optimize order $q$ over $\mathcal{M}$}
    \For{each $q_c \in \mathcal{M}$}
        \State $J_c \leftarrow J(q_c,\eta_{\mathrm{nuc}},\eta_{\mathrm{post}})$
        \If{$J_c < J^\star$} 
            $(q,J^\star) \leftarrow (q_c,J_c)$
        \EndIf
    \EndFor 

    \Statex \hspace{0.5em}\emph{(2) Optimize $\eta_{\mathrm{nuc}}$ via step search}
    \For{each $s \in \mathcal{S}$, $d \in \{+1,-1\}$}
        \State $\eta_c \leftarrow \mathrm{clip}(\eta_{\mathrm{nuc}} + d\,s,\eta_{\min},\eta_{\max})$
        \If{$J(q,\eta_c,\eta_{\mathrm{post}}) < J^\star$}
            $(\eta_{\mathrm{nuc}},J^\star) \leftarrow (\eta_c, J(q,\eta_c,\eta_{\mathrm{post}}))$; \textbf{Break}
        \EndIf
    \EndFor

    \Statex \hspace{0.5em}\emph{(3) Optimize $\eta_{\mathrm{post}}$ via step search}
    \For{each $s \in \mathcal{S}$, $d \in \{+1,-1\}$}
        \State $\eta_c \leftarrow \mathrm{clip}(\eta_{\mathrm{post}} + d\,s,\eta_{\min},\eta_{\max})$
        \If{$J(q,\eta_{\mathrm{nuc}},\eta_c) < J^\star$}
            $(\eta_{\mathrm{post}},J^\star) \leftarrow (\eta_c, J(q,\eta_{\mathrm{nuc}},\eta_c))$; \textbf{Break}
        \EndIf
    \EndFor

    \If{$J_{\mathrm{prev}} - J^\star < \tau_{\mathrm{stop}}$} \textbf{Break} \EndIf
\EndFor

\State \Return $(q,\eta_{\mathrm{nuc}},\eta_{\mathrm{post}},J^\star)$
\end{algorithmic}
\end{algorithm}

% \subsection{Monte Carlo EM analysis}
% In our Monte Carlo analysis, we follow the work in~\cite{Chatterjee:ICCAD'16,Najm:IRPS2019}, where, EM diffusivity is treated as a random variables. Since our method is general Monte Carlo method, we can consider any variational sources. For each wire, we first performance sensitivity analysis to determine the best order and shift time for nucleation and post-voids. The requirement is less than 1\% error for both nucleation time and segment wire resistance change, which is very achievable for order 4-6 using ExtRaKryloveEM solver. 

\section{EM Stress Analysis via Exponential Integration With Rational Krylov Methods}
\label{sec:rakrylov_ei_method}
In this section, we present the exponential-integration method based on rational Krylov methods, called {\it EiRaKrylovEM}.

We first review the exponential-integration framework for solving linear ODEs, followed by the rational Krylov subspace approximation of the matrix-exponential operator. We then present the residual error estimation.
To begin, for the constant-current case and using the corresponding standard-form system, we rewrite the linear ordinary differential equation from \eqref{eq:lti_em} as follows:
\begin{equation}
\frac{d\boldsymbol{\sigma}(t)}{dt} = A\,\boldsymbol{\sigma}(t) + \boldsymbol{b},
\qquad
\boldsymbol{b} = B\,\boldsymbol{j},
\label{eq:lti_em2}
\end{equation}
assuming constant input $\boldsymbol{j}$ throughout this derivation and initial
condition $\boldsymbol{\sigma}(0)=\boldsymbol{\sigma}_0$.
Define
\begin{equation}
\boldsymbol{f} = A^{-1}\boldsymbol{b},
\qquad
\boldsymbol{v}_0 = \boldsymbol{\sigma}_0 + \boldsymbol{f}.
\end{equation}
The closed-form solution for \eqref{eq:lti_em2} is
\begin{align}
\boldsymbol{\sigma}(t)
&= e^{tA}\bigl(\boldsymbol{\sigma}_0 + A^{-1}\boldsymbol{b}\bigr)
   - A^{-1}\boldsymbol{b}.\\
&= e^{tA}\boldsymbol{v}_0 - \boldsymbol{f}
\label{eq:sigma_ei_closed}
\end{align}
Computing the matrix exponential $e^{tA}\boldsymbol{v}_0$ is generally very
expensive. The key idea of the EI-based method is to approximate it using a
rational Krylov subspace method.

\subsection{Rational Krylov Approximation for the Matrix Exponential}
Specifically, the key computing step is to approximate
$e^{tA}\boldsymbol{v}_0$ in \eqref{eq:sigma_ei_closed} using a rational Krylov
subspace method. There are several Krylov subspaces we can select, such as
standard $\mathcal{K}_q(A, \boldsymbol{v}_0)$, inverse
$\mathcal{K}_q(A^{-1}, \boldsymbol{v}_0)$, and rational Krylov subspaces
$\mathcal{K}_q\!\big((A - s_0 I)^{-1}, \boldsymbol{v}_0\big)$~\cite{Bochev:short_guide2012}. 

It was shown that the rational Krylov subspace can provide much better
approximation than the standard and inverse Krylov subspaces in the EI framework
for IR drop analysis~\cite{Zhuang:TCAD2016}. However, the shift or expansion
point $s_0$ in this method does not correlate well to the accuracy of the
circuit being analyzed. We further note that the work in~\cite{Stoikos:SMACD23} used
$\mathcal{K}_q\!\big((A^{-1} - s_0 I)^{-1}, \boldsymbol{v}_0\big)$, instead of
the typical rational Krylov subspace
$\mathcal{K}_q\!\big((A - s_0 I)^{-1}, \boldsymbol{v}_0\big)$.

In this paper, we show that the standard rational Krylov subspace
$\mathcal{K}_q\!\big((A - s_0 I)^{-1}, \boldsymbol{v}_0\big)$ is a
good choice for the EM stress analysis problem as the shift $s_0$ can be
aligned with times of interest such as the nucleation time or steady-state time,
which are wire- and application-specific in EM analysis. 

For rational Krylov subspace
$\mathcal{K}_q\!\big((A - s_0 I)^{-1}, \boldsymbol{v}_0\big)$, let $H_q$ denote
the resulting reduced (square) Hessenberg matrix. Let
$V_q = [\boldsymbol{v}_1, ..., \boldsymbol{v}_q]$ be the orthogonal basis of the
rational Krylov subspace, which satisfies the Arnoldi
decomposition~\eqref{eq:arnoldi_decomposition}. 
% \begin{equation}
%     \big((A - s_0 I)^{-1}V_q = V_q H_q + h_{q+1},q v_{q+1}e^T_q
%     \label{eq:arnoldi_decomp}
% \end{equation}
Then the mapped reduced operator is defined as
\begin{equation}
\widehat{H}_q = H_q^{-1} + s_0 I_q,
\label{eq:H_head}
\end{equation}
which corresponds to the identity
\begin{equation}
A = s_0 I + \bigl((A-s_0 I)^{-1}\bigr)^{-1}
\end{equation}
in the reduced space. Let $V_q\in \mathbb{R}^{n\times q}$ be the Krylov basis,
$\beta = \|\boldsymbol{v}_0\|_2$, and $e_1 = [1,0,\dots,0]^T \in \mathbb{R}^q$.
The exponential integration (EI) approximation is given by~\cite{Bochev:short_guide2012}:
\begin{equation}
e^{tA}\boldsymbol{v}_0 \;\approx\; \beta\,V_q\,e^{t\widehat{H}_q}e_1.
\end{equation}

Algorithm~\ref{alg:rk_expm_last_h_v} constructs a rational Krylov subspace using repeated applications of $(A-s_0 I)^{-1}$ and returns the reduced Hessenberg matrix. It returns not only $V_q$ and $H_q$, but also
$\beta$, $h_{k+1,k}$, and $v_{k+1}$, which are used later for residual-error computation. Note that $q=k$ if the process completes to order $q$. 

\begin{algorithm}[t]
\caption{Rational Krylov Subspace Reduction With Residual Information}
\label{alg:rk_expm_last_h_v}
\begin{algorithmic}[1]
\Require
Matrix $A \in \mathbb{R}^{n \times n}$, Vector $v \in \mathbb{R}^n$, 
Maximum Krylov Dimension $q$, Expansion Point $s_0$
\Ensure
Basis $V_m = [v_1,\dots,v_k]$, Hessenberg Matrix $H_m \in \mathbb{R}^{k \times k}$,
$\beta = \|v\|_2$, Last-Row Coefficients $h_{k+1,k}$, Next Basis Vector $v_{k+1}$

\State $\beta \leftarrow \|v\|_2$ and $v_1 \leftarrow v / \beta$
% \If{$\beta = 0$}
%     \State \textbf{Error}: Zero Starting Vector
% \EndIf
%\State $v_1 \leftarrow v / \beta$

% \If{Expansion Point $s_0$ Is Not Provided}
%     \State $s_0 \leftarrow \mathrm{mean}(\mathrm{diag}(A))$
% \EndIf

%\State Initialize $V \in \mathbb{R}^{n \times (m+1)}$, $H \in \mathbb{R}^{(m+1) \times m}$
\State $V(:,1) \leftarrow v_1$

\State Form $M \leftarrow A - s_0 I$
\State Compute LU Factorization $M = LU$

\State $k \leftarrow 0$ and $v_{k+1} \leftarrow \emptyset$

\For{$j = 1$ to $q$}
    \State Solve $w \leftarrow M^{-1} V(:,j)$ Using LU Factors
    \For{$i = 1$ to $j$}
        \State $h_{i,j} \leftarrow V(:,i)^{T} w$
        \State $w \leftarrow w - h_{i,j} V(:,i)$
    \EndFor
    \State $h_{j+1,j} \leftarrow \|w\|_2$
    \If{$h_{j+1,j} < \varepsilon$}
        \State $k \leftarrow j$
        \State $v_{k+1} \leftarrow \emptyset$ \Comment{Happy breakdown}
        \State \textbf{Break}
    \EndIf
    \State $V(:,j+1) \leftarrow w / h_{j+1,j}$
    \State $v_{k+1} \leftarrow V(:,j+1)$
    \State $k \leftarrow j$
\EndFor

% \If{$k = 0$}
%     \State $k \leftarrow 1$
% \EndIf

\State $V_q \leftarrow V(:,1:k)$
\State $H_q \leftarrow H(1:k,1:k)$
\State $h_{k+1,k} \leftarrow H(k+1,1:k)$

\State \Return $V_q, H_q, \beta, h_{k+1,k}, v_{k+1}$
\end{algorithmic}
\end{algorithm}

After the $\widehat{H}_q$ is calculated in \eqref{eq:H_head}, the stress at
time $t$ can be computed as
\begin{equation}
\boldsymbol{\sigma}_q(t)=\beta\,V_q\,e^{t\widehat{H}_q}e_1-\boldsymbol{f}.
\end{equation}
For $t=0$, the initial condition
$\boldsymbol{\sigma}(0)=\boldsymbol{\sigma}_0$ is applied explicitly. The whole
time-domain analysis based on EI with rational Krylov subspace is summarized in
Algorithm~\ref{alg:ei_time_series}. 

\subsection{Residual Error Estimation}
One important advantage of the EI-Krylov-based method is that it provides a residual-error estimate in the time domain. 
Let the reduced state be
\begin{equation}
\boldsymbol{z}_q(t) = e^{t\widehat{H}_q}e_1.
\end{equation}
The residual vector is approximated as~\cite{Bochev:short_guide2012}
\begin{equation}
\boldsymbol{r}_q(t)
\;\approx\;
\beta\;\langle h_{q+1,q}\boldsymbol{e}_q,\,\boldsymbol{z}_q(t)\rangle\; \boldsymbol{v}_{q+1},
\end{equation}
where $\langle\cdot,\cdot\rangle$ denotes the standard inner product. The absolute norm is then
\begin{equation}
\|\boldsymbol{r}_q(t)\|_2
=
\left\|
\beta\,\bigl(h_{q+1,q}\,\boldsymbol{z}_q(t)\bigr)\,\boldsymbol{v}_{q+1}
\right\|_2,
\label{eq:time_error}
\end{equation}
and the relative residual error is defined as
\begin{equation}
r_{rel}(t)
=
\frac{\|\boldsymbol{r}_q(t)\|_2}
{\max\bigl(\|\boldsymbol{\sigma}_q(t)\|_2,\;10^{-30}\bigr)}.
\label{eq:abs_time_error}
\end{equation}

For the {\it EiRaKrylovEM} method, we use this residual estimate in the results section and to guide the selection of reduction order and shift time for both the nucleation and post-void phases.

To determine the best reduction order and shift times for the nucleation and post-void phases, we again apply the coordinate-descent Algorithm~\ref{alg:deep_descent_krylov} presented in the previous section to minimize the combined error metric based on the nucleation time and the resistance change of the nucleated wire segment.

Note that in the previous discussion we assume the input current density
$\boldsymbol{j}$ is constant during the entire simulation time. However, this
is one major drawback of the {\it EiRaKrylovEM} method compared with
the frequency-domain {\it ExtRaKrylovEM}. If the current waveform becomes
time-varying~\cite{Sukharev:JAP2015}, the Krylov subspace has to be rebuilt for
each time interval in which $\boldsymbol{j}(t)$ changes, which significantly
diminishes the efficiency advantage of the method.

% \subsection{Zero-Mean Projection for the Nucleation Phase}

% For the nucleation phase, a volume-weighted zero-mean constraint is enforced:
% \begin{equation}
% \bar{\sigma}(t)
% =
% \frac{\sum_i \sigma_i(t)\,\mathrm{vol}_i}
% {\sum_i \mathrm{vol}_i},
% \end{equation}
% followed by the projection
% \begin{equation}
% \sigma(t) \leftarrow \sigma(t) - \bar{\sigma}(t)\mathbf{1}.
% \end{equation}

\begin{algorithm}[t]
\caption{Exponential Integration Using Single-Shift Rational Krylov}
\label{alg:ei_time_series}
\begin{algorithmic}[1]
\Require $\mathbf{A}, \mathbf{B}, \boldsymbol{j}, \boldsymbol{\sigma}_0, \{t_k\}, q, s_0$
\Ensure Stress trajectory $\{\boldsymbol{\sigma}_q(t_k)\}$

\State $\mathbf{b}\leftarrow \mathbf{B}\boldsymbol{j}$
\State $\mathbf{f}\leftarrow \mathbf{A}^{-1}\mathbf{b}$
\State $\mathbf{v}_0\leftarrow \boldsymbol{\sigma}_0+\mathbf{f}$
\State $\beta\leftarrow\|\mathbf{v}_0\|_2$

\State Construct rational Krylov basis $\mathbf{V}_q,\mathbf{H}_q,
\mathbf{v}_{q+1},\mathbf{h}_{q+1,1:q}$ using $(\mathbf{A}-s_0\mathbf{I})^{-1}$

\State $\widehat{\mathbf{H}}_q\leftarrow\mathbf{H}_q^{-1}+s_0\mathbf{I}$

\For{each $t_k$}
    \If{$t_k=0$}
        \State $\boldsymbol{\sigma}_q(t_k)\leftarrow\boldsymbol{\sigma}_0$
    \Else
        \State $\mathbf{z}\leftarrow e^{t_k\widehat{\mathbf{H}}_q}\mathbf{e}_1$
        \State $\boldsymbol{\sigma}_q(t_k)\leftarrow \beta\mathbf{V}_q\mathbf{z}-\mathbf{f}$
        \State Compute residual using \eqref{eq:time_error} and \eqref{eq:abs_time_error}
    \EndIf
\EndFor

\Return $\{\boldsymbol{\sigma}_q(t_k)\}$
\end{algorithmic}
\end{algorithm}

\section{Experimental Results}
\label{sec:results}

% ==================================== Clean Version ===============================================
\subsection{Experimental Setup}

% The proposed \textit{\textbf{BPINN-EM-Post}} framework is fully developed using Python/PyTorch. 
% The training and test procedures are conducted on a Linux server equipped with two Intel 22-core E5-2699 CPUs, 320 GB of memory, and an Nvidia TITAN RTX GPU.
% A three-layer MLP is implemented in BPINN flux predictor.
% with layer structure [I6-FC50-FC50-O3]
Our numerical experiments are conducted on an iMac with an Apple M3 chip (8-core CPU and 32GB memory).
% To validate the accuracy and performance of our proposed \textit{ExtRaKrylovEM}, we utilized power grids on two categories of designs:
% 1) Synthesized multi-segment trees with different number of segments to reflect practical power grid cases.
% We utilize trees with 100, 200, 500, 1000, 2000 segments randomly generated with various segment geometry and current density, following multi-segment tree generation algorithm in~\cite{Jin:DAC'21}.
% 2) To demonstrate the practicality of our proposed method, we extract the three longest trees in the power grids of two real designs ARM-Cortex~\cite{AMDCortexM0} and ASE IP from OpenROAD~\cite{OpenROAD_DAc19} considering the vulnerability of long trees to EM.
% Accuracy of stress prediction and nucleation time estimation are employed to evaluate the performance of \textit{ExtRaKrylovEM}, and analysis runtime is used to demonstrate the speedup compared to prior arts~\cite{SunYu:TDMR'20}.

The proposed \textit{ExtRaKrylovEM} and \textit{EiRaKrylovEM} frameworks are fully implemented in Python v3.11. 
We also implement FastEM~\cite{CookSun:TVLSI'18} and use the implementation from~\cite{SunYu:TDMR'20} as the FDM solver, which serves as the baseline for accuracy and performance comparison.
To assess the accuracy and efficiency of \textit{ExtRaKrylovEM} and \textit{EiRaKrylovEM} solvers, we evaluate them on two categories of power-grid designs.
(1) Synthetic multi-segment trees: We generate trees containing 100, 200, 500, 1000, and 2000 segments with random geometries and current densities, following the multi-segment interconnect tree generation algorithm in~\cite{Jin:DAC'21} to cover designs of varying sizes. 
These cases emulate practical power-grid structures with varying complexity, including segment geometry and current density.
(2) Real design benchmarks: To further demonstrate the practicality of our method, we extract the three longest interconnect trees from the power grids of industrial designs, including ARM-Cortex~\cite{AMDCortexM0} and the IBM Power Grid Benchmarks~\cite{IBM_PG_Bench},
%the AES IP from OpenROAD~\cite{OpenROAD_DAc19}, 
because long trees are generally more vulnerable to EM. These circuits are synthesized in 32nm technology. Stress-prediction accuracy and nucleation-time $t_{\text{nuc}}$ estimation are used to evaluate the fidelity of our proposed methods, while the total runtime is reported to demonstrate the efficiency improvement over prior work~\cite{SunYu:TDMR'20}.

\subsection{Sensitivity Analysis for Different Configurations}
First, we perform a simple sensitivity analysis to see how the model order $q$ and shift factors $\{\eta_{\text{nuc}}, \eta_{\text{post}}\}$ affect the accuracy of the proposed methods.

We conduct a case study on a synthesized 500-segment interconnect tree representative of a typical block-level power grid.
Table~\ref{tab:Sensitivity_overall} summarizes the errors obtained across different model orders for the synthesized 500-segment interconnect tree. 
% \usepackage{multirow}
\renewcommand{\arraystretch}{1.2}

\begin{table}[htp]
    \centering
    \caption{\small Sensitivity Analysis results for our proposed (a) \textit{ExtRaKrylovEM} and (b) \textit{EiRaKrylovEM} on the 500-segment wire}
    \label{tab:Sensitivity_overall}
    \begin{subtable}[t]{0.49\textwidth}
        \centering
        \caption{\textit{ExtRaKrylovEM}}
        \vspace{-5pt}
        \label{tab:Sensitivity_ExtRaKrylovEM}
        \small
        \resizebox{\linewidth}{!}{
        \begin{tabular}{c|cc|cc|cc|cc}
            \hline\hline
            \begin{tabular}[c]{@{}c@{}}Model\\Order $q$\end{tabular}
            & \multicolumn{2}{c|}{$q=3$}
            & \multicolumn{2}{c|}{$q=4$}
            & \multicolumn{2}{c|}{$q=5$}
            & \multicolumn{2}{c}{$q=6$} \\
            \hline
            Phase & $\eta_{\text{Nuc}}$ & $\eta_{\text{Post}}$ & $\eta_{\text{Nuc}}$ & $\eta_{\text{Post}}$ & $\eta_{\text{Nuc}}$ & $\eta_{\text{Post}}$ & $\eta_{\text{Nuc}}$ & $\eta_{\text{Post}}$ \\
            \hline
            $s_f$  & 15.5 & 0.6 & 3.5 & 20  & 0.2 & 3.5 & 0.2 & 4 \\
            Error $\epsilon$  & 0.06\% & 0.00\% & 0.01\% & 0.00\% & 0.00\% & 0.00\% & 0.00\% & 0.00\% \\
            \hline\hline
            
        \end{tabular}
        }
        \vspace{5pt}
    \end{subtable}
    \hfill
    \begin{subtable}[t]{0.49\textwidth}
        \centering
        \caption{\textit{EiRaKrylovEM}}
        \vspace{-5pt}
        \label{tab:Sensitivity_EiRaKrylovEM}
        \small
        \resizebox{\linewidth}{!}{%
        \begin{tabular}{c|cc|cc|cc|cc}
        \hline\hline
        \begin{tabular}[c]{@{}c@{}}Model\\Order $q$\end{tabular}
        & \multicolumn{2}{c|}{$q=3$}
        & \multicolumn{2}{c|}{$q=4$}
        & \multicolumn{2}{c|}{$q=5$}
        & \multicolumn{2}{c}{$q=6$} \\
        \hline
        Phase & $\eta_{\text{Nuc}}$ & $\eta_{\text{Post}}$ & $\eta_{\text{Nuc}}$ & $\eta_{\text{Post}}$ & $\eta_{\text{Nuc}}$ & $\eta_{\text{Post}}$ & $\eta_{\text{Nuc}}$ & $\eta_{\text{Post}}$ \\
        \hline
        $s_f$  & 0.3 & 20   & 6.5 & 19.5 & 0.1 & 18.5 & 0.4 & 0.8 \\
        Error $\epsilon$ & 1.11\% & 0.02\% & 0.02\% & 0.00\% & 0.03\% & 0.00\% & 0.01\% & 0.00\% \\
        \hline\hline
        \end{tabular}
        }
    \end{subtable}
\end{table}
% Our results indicate that as the model order $q$ increases, the prediction error rapidly decreases and then stabilizes, indicating convergence of the higher-order sensitivity terms.
% In particular, the accuracy gain saturates at $q=5$, beyond which further increasing the order provides negligible improvement.

% \usepackage{multirow}
\begin{table*}[htp]
\centering
\caption{\small Performance Comparison among the proposed \textit{ExtRaKrylovEM} and existing FDM~\cite{SunYu:TDMR'20} and FastEM~\cite{CookSun:TVLSI'18} methods with 100 simulation time steps.}
\resizebox{\linewidth}{!}{
\begin{tabular}{c|c|cccccccccccc}
\hline\hline
\multirow{4}{*}{Design}                                        & \multicolumn{1}{c|}{\multirow{4}{*}{\begin{tabular}[c]{@{}l@{}}\# of\\ Segments\end{tabular}}} & \multicolumn{12}{c}{Methods}                                                                                                                                                                                                                                                                                                      \\ \cline{3-14} 
\multicolumn{1}{l|}{}                                                               & \multicolumn{1}{l|}{}                                                                          & \multicolumn{1}{c|}{FDM~\cite{SunYu:TDMR'20}}         & \multicolumn{4}{c|}{Krylov-based FastEM~\cite{CookSun:TVLSI'18}}                                & \multicolumn{7}{c}{Proposed \textit{ExtRaKrylovEM}}                                                                                                                                                                           \\ \cline{3-14} 
\multicolumn{1}{l|}{}                                                               & \multicolumn{1}{l|}{}                                                                          & \multicolumn{1}{c|}{Runtime (s)} & \begin{tabular}[c]{@{}c@{}}Model \\ Order\end{tabular}  & $\epsilon_{\text{nuc}}$ & $\epsilon_{\text{post}}$ & \multicolumn{1}{c|}{Runtime (s)} & Config$^*$          & $\epsilon_{\text{nuc}}$ & $\epsilon_{\text{post}}$  & Total Error & \multicolumn{1}{c|}{Runtime (s)} & \begin{tabular}[c]{@{}c@{}}Speedup \\ to FDM\end{tabular} & \begin{tabular}[c]{@{}c@{}}Speedup to \\ FastEM\end{tabular} \\ \hline
\multirow{5}{*}{\begin{tabular}[c]{@{}c@{}}Synthesized\\ Power Grids\end{tabular}}  
& 100                                                                                            & \multicolumn{1}{c|}{2.01}        & 50    & 64.75\%   & 0.00\%     & \multicolumn{1}{c|}{0.15}        & \{4, 1.0, 1.0\} & 0.01\%    & 0.00\%     & 0.01\%      & \multicolumn{1}{c|}{0.035}       & 57.22$\times$                                                     & 4.29$\times$                                                         \\
& 200                                                                                            & \multicolumn{1}{c|}{13.92}       & 50          & 3.86\%    & 0.00\%     & \multicolumn{1}{c|}{0.42}        & \{4, 1.1, 1.0\} & 0.06\%    & 0.00\%     & 0.06\%      & \multicolumn{1}{c|}{0.171}       & 81.30$\times$                                                     & 2.43$\times$                                                         \\
                                                                                    & 500                                                                                            & \multicolumn{1}{c|}{132.31}      & 50          & 0.25\%    & 0.00\%     & \multicolumn{1}{c|}{2.23}        & \{6, 0.5, 1.0\} & 0.02\%    & 0.00\%     & 0.02\%      & \multicolumn{1}{c|}{1.172}       & 112.94$\times$                                                    & 1.90$\times$                                                         \\
                                                                                    & 1000                                                                                           & \multicolumn{1}{c|}{914.39}      & 50          & 12.65\%   & 0.00\%     & \multicolumn{1}{c|}{10.04}       & \{6, 0.5, 1.0\} & 0.00\%    & 0.00\%     & 0.00\%      & \multicolumn{1}{c|}{7.131}       & 128.22$\times$                                                    & 1.41$\times$                                                         \\
                                                                                    & 2000                                                                                           & \multicolumn{1}{c|}{2124.84}     & 50          & 12.36\%   & 0.01\%     & \multicolumn{1}{c|}{83.62}       & \{4, 1.0, 1.0\} & 0.13\%    & 0.00\%     & 0.13\%      & \multicolumn{1}{c|}{60.223}      & 35.28$\times$                                                     & 1.39$\times$                                                         \\ \hline
ARM Cortex~\cite{AMDCortexM0}                                                                       & 33                                                                                             & \multicolumn{1}{c|}{0.06}        & 50          & 0.00\%    & 0.00\%     & \multicolumn{1}{c|}{0.01}        & \{6, 2.9, 1.0\} & 0.00\%    & 0.00\%     & 0.00\%      & \multicolumn{1}{c|}{0.003}       & 18.74$\times$                                                     & 3.71$\times$                                                         \\ \hline
\multirow{4}{*}{\begin{tabular}[c]{@{}c@{}}IBM Power \\ Grid Benmarks~\cite{IBM_PG_Bench} \end{tabular}} & 236                                                                                            & \multicolumn{1}{c|}{21.07}       & 50          & 1.29\%    & 0.00\%     & \multicolumn{1}{c|}{0.52}        & \{6, 9.0, 1.0\} & 0.00\%    & 0.00\%     & 0.00\%      & \multicolumn{1}{c|}{0.213}       & 98.85$\times$                                                     & 2.43$\times$                     \\
                                                                                    & 572                                                                                            & \multicolumn{1}{c|}{286.24}      & 50          & 0.97\%    & 0.00\%     & \multicolumn{1}{c|}{3.64}        & \{6, 0.3, 1.0\} & 0.02\%    & 0.00\%     & 0.02\%      & \multicolumn{1}{c|}{1.514}       & 189.11$\times$                                                    & 2.40$\times$                                                         \\
                                                                                    & 1275                                                                                           & \multicolumn{1}{c|}{489.78}      & 50          & 4.64\%    & 0.00\%     & \multicolumn{1}{c|}{21.07}       & \{4, 3.0, 1.0\} & 0.00\%    & 0.00\%     & 0.00\%      & \multicolumn{1}{c|}{13.468}      & 36.37$\times$                                                     & 1.56$\times$                                                         \\
                                                                                    & 2403                                                                                           & \multicolumn{1}{c|}{3193.55}     & 50          & 4.93\%    & 0.00\%     & \multicolumn{1}{c|}{112.08}      & \{5, 0.2, 1.0\} & 0.00\%    & 0.00\%     & 0.00\%      & \multicolumn{1}{c|}{92.268}      & 34.61$\times$                                                  & 1.22$\times$                                                         \\ \hline
    \textbf{Average}                                                                             & \textbf{832}                                                                                          & \multicolumn{1}{c|}{\textbf{717.82}}      & \textbf{50}          & \textbf{5.94\%}    & \textbf{0.00\%}     & \multicolumn{1}{c|}{\textbf{23.38}}       & -               & \textbf{0.02\%}    & \textbf{0.00\%}     & \textbf{0.02\%}      & \multicolumn{1}{c|}{\textbf{17.620}}      & \textbf{79.26$\times$}                                                    & \textbf{2.27$\times$}                                                        \\ \hline\hline
\end{tabular}
}
\label{tab:main_result_extra}
%\vspace{pt}
{\footnotesize $^{*}$Config = \{Krylov order, nucleation shift factor, post-void shift factor\}.}
\end{table*}

\begin{table*}[t]
\centering
\caption{\small Performance comparison between FDM~\cite{SunYu:TDMR'20} and the proposed \textit{EiRaKrylovEM} using 100 simulation time steps.}
% >>>>>>> 882cce3 (add changes)
\label{tab:main_result_ei}
\setlength{\tabcolsep}{4pt}
\begin{tabular}{c|c|c|c|ccc|c|c}
\hline\hline
Design 
& \begin{tabular}[c]{@{}c@{}}\# of\\Segments\end{tabular}
& \begin{tabular}[c]{@{}c@{}}FDM\\Runtime (s)\end{tabular}
& Config$^{*}$
& $\epsilon_{\text{nuc}}$
& $\epsilon_{\text{post}}$
& Total Error
& \begin{tabular}[c]{@{}c@{}}EiRaKrylovEM\\Runtime (s)\end{tabular}
& \begin{tabular}[c]{@{}c@{}}Speedup\\to FDM\end{tabular}
\\
\hline

\multirow{5}{*}{\begin{tabular}[c]{@{}c@{}}Synthesized\\Power Grids\end{tabular}}
& 100  & 2.01    & \{6, 3.4, 11.0\} & 0.02\% & 0.00\% & 0.02\% & 0.040   & 45.75$\times$ \\
& 200  & 13.92   & \{4, 0.2, 4.2\}  & 0.03\% & 0.00\% & 0.03\% & 0.270   & 51.19$\times$ \\
& 500  & 132.31  & \{6, 0.4, 7.0\}  & 0.01\% & 0.00\% & 0.01\% & 1.720   & 77.11$\times$ \\
& 1000 & 914.39  & \{5, 3.0, 2.0\}  & 0.01\% & 0.00\% & 0.01\% & 11.175  & 81.82$\times$ \\
& 2000 & 2124.84 & \{6, 0.6, 0.5\}  & 0.08\% & 0.00\% & 0.08\% & 113.526 & 18.72$\times$ \\
\hline

ARM Cortex-M0~\cite{AMDCortexM0}
& 33 & 0.06 & \{5, 1.2, 2.5\} & 0.01\% & 0.00\% & 0.01\% & 0.019 & 3.41$\times$ \\
\hline

\multirow{4}{*}{\begin{tabular}[c]{@{}c@{}}IBM Power Grid\\Benchmarks~\cite{IBM_PG_Bench}\end{tabular}}
& 236  & 21.07   & \{6, 0.5, 3.5\} & 0.12\% & 0.00\% & 0.12\% & 0.367   & 57.42$\times$ \\
& 572  & 286.24  & \{6, 0.8, 1.0\} & 0.04\% & 0.00\% & 0.04\% & 2.338   & 122.43$\times$ \\
& 1275 & 489.78  & \{5, 3.0, 1.0\} & 0.01\% & 0.00\% & 0.01\% & 23.234  & 21.08$\times$ \\
& 2403 & 3193.55 & \{4, 1.2, 5.0\} & 0.06\% & 0.00\% & 0.06\% & 164.238 & 19.44$\times$ \\ \hline

\textbf{Average}  & \textbf{832}                                                                     & \multicolumn{1}{c|}{\textbf{717.82}}        & -               & \textbf{0.04\%}    &\textbf{ 0.00\% }    & \textbf{0.04\% }     & \multicolumn{1}{c|}{\textbf{31.69}}       & \textbf{49.84$\times$}                                                               \\ \hline
\hline
\end{tabular}

\vspace{2pt}
{\footnotesize $^{*}$Config = \{Krylov order, nucleation shift factor, post-void shift factor\}.}
\end{table*}

We perform a coarse sweep over a set of candidate shift values $s_f$ to identify the optimal configuration of model order $q$ and shift factors $\{\eta_{\text{nuc}}, \eta_{\text{post}}\}$ that minimizes the analysis error for both nucleation time and resistance change. As shown in Table~\ref{tab:Sensitivity_ExtRaKrylovEM}, the sensitivity analysis identifies the optimal configuration $\{q^*=5,\; \eta_{\text{nuc}}^* = 0.2,\; \eta_{\text{post}}^* = 3.5\}$, which reduces the reported errors for both nucleation time and resistance change to 0.00\%. In contrast, the baseline solver without sensitivity correction exhibits a nucleation-time error exceeding 13.6\%.

\subsection{Accuracy and Performance of \textit{ExtRaKrylovEM} and \textit{EiRaKrylovEM} for Deterministic Analysis}

To assess the accuracy and computational efficiency of the proposed \textit{ExtRaKrylovEM} and \textit{EiRaKrylovEM} methods relative to existing approaches, Table~\ref{tab:main_result_extra} and Table~\ref{tab:main_result_ei} report performance comparisons across a set of wire benchmarks. 
The evaluated methods include our \textit{ExtRaKrylovEM}, the finite-difference (FDM) solver~\cite{SunYu:TDMR'20}, and the Krylov-based FastEM framework~\cite{CookSun:TVLSI'18}.

% For consistency, we set the critical stress threshold to $\sigma_{\mathrm{crit}} = 10^{8}$~Pa for trees with fewer than 500 segments and $\sigma_{\mathrm{crit}} = 10^{9}$~Pa for larger structures.

%\input{tables/tab_MC}

% \input{tables/tab_single_tree_comparison}

% \input{tables/tab_single_tree_comparison}
For consistency, we set the critical stress threshold to $\sigma_{\mathrm{crit}} = 10^{8}$~Pa for all analyses. For the standard Krylov-based FastEM method, we use a reduction order of 50 to ensure a fair comparison, as our methods also employ a nearly fixed order (4--6). Table~\ref{tab:main_result_extra} summarizes the performance comparison among the proposed \textit{ExtRaKrylovEM}, the FDM solver~\cite{SunYu:TDMR'20}, and FastEM~\cite{CookSun:TVLSI'18} using 100 simulation time steps.

\begin{figure}[h]
    \centering
    \includegraphics[width=\linewidth]{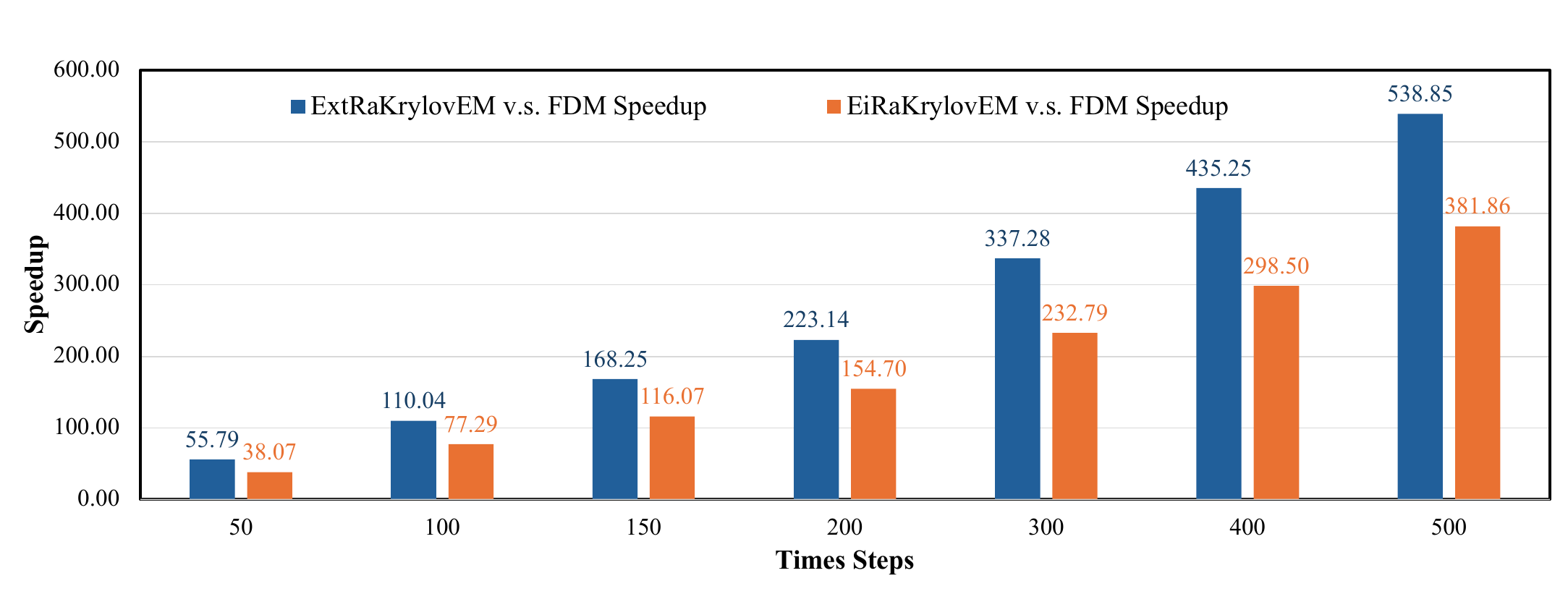}
    \vspace{-10pt}
    \caption{\small Comparison between our proposed \textit{ExtRaKrylovEM} and \textit{EiRaKrylovEM} with Back Euler based FDM under different analysis time steps for 500 segment wire.}
    \vspace{-5pt}
    \label{fig:timestep}
\end{figure}

From Table~\ref{tab:main_result_extra}, we observe that the proposed \textit{ExtRaKrylovEM} achieves approximately 34--189$\times$ speedup (80$\times$ on average) over the FDM-based solver for 100 time steps. The speedup also scales with the number of required time steps, as further illustrated in Fig.~\ref{fig:timestep}.

While \textit{ExtRaKrylovEM} delivers only about $2.3\times$ speedup compared to FastEM, the key distinction lies in accuracy. FastEM exhibits significant errors in nucleation-time estimation, whereas our method attains near-zero error (less than 0.1\% on average). As demonstrated in the motivating example in Section~\ref{sec:review}, such errors can lead to incorrect identification of the nucleation location (cathode node), which is unacceptable for reliable EM analysis. Consequently, a meaningful speedup comparison is only applicable against the FDM method, which serves as our accuracy baseline.

We also observe that all methods estimate the resistance change with nearly zero error. This is because the resistance change is mainly determined by the void size at steady state, which lies on a very long time scale where both the standard Krylov method (expanded at infinite time) and the rational Krylov method provide good accuracy.

Table~\ref{tab:main_result_ei} presents the performance comparison between the proposed \textit{EiRaKrylovEM} and the FDM solver~\cite{SunYu:TDMR'20} using 100 simulation time steps. As shown, \textit{EiRaKrylovEM} achieves approximately 18--122$\times$ acceleration (50$\times$ on average) over FDM while maintaining similar accuracy with sub-0.1\% error in both nucleation time and resistance change. Overall, both proposed methods, with very low reduction orders (3--6), achieve significant speedup over the FDM solver with negligible accuracy loss, making them highly suitable for stochastic EM analysis and EM-aware optimization.

% Fig.~\ref{fig:stress_comp_wire1_10seg} shows examples of the stress maps produced by \textit{ExtRaKrylovEM} under the optimal shifting configurations  on a 10-segment wire (order = 4, nucleation time error is 0.01\% and resistance change error is 0.046\%) (at bottom). 
% Results obtained from FDM-based solver~\cite{SunYu:TDMR'20} are also shown at the top. We show three time points, the nucleation time (left one), the middle of post-void (middle), and end of post-void phase (right one).

Furthermore, we demonstrate that the speedups of both \textit{ExtRaKrylovEM} and \textit{EiRaKrylovEM} over FDM scale linearly with the number of analysis time steps, highlighting the strong scalability of the proposed methods.
Fig.~\ref{fig:timestep} shows the runtime comparison of the proposed \textit{ExtRaKrylovEM} and \textit{EiRaKrylovEM} against the FDM solver~\cite{SunYu:TDMR'20} on the 500-segment interconnect under varying numbers of analysis time steps. The optimal configuration for \textit{ExtRaKrylovEM} is $\{q^*, \eta_{\text{nuc}}^*, \eta_{\text{post}}^*\} = \{6,\, 0.5,\, 1.0\}$, and for \textit{EiRaKrylovEM} is $\{q^*, \eta_{\text{nuc}}^*, \eta_{\text{post}}^*\} = \{6,\, 0.4,\, 7.0\}$. 

As shown in the figure, the speedup over FDM increases linearly with the number of time steps for both methods, reaching up to $300$--$500\times$ with 500 time steps. This strong scalability arises because the FDM solver must perform time stepping at each time point, whereas both Krylov-based methods only need to construct the reduced-order model once and can subsequently evaluate it at any time point with negligible additional cost due to the small model order. This characteristic makes them particularly advantageous for applications requiring fine temporal resolution or extended aging analysis. We also observe that \textit{ExtRaKrylovEM} is faster than \textit{EiRaKrylovEM} for the same number of time steps, while both methods maintain near-zero error in nucleation time and resistance change, demonstrating that scalability does not come at the cost of accuracy. We do not compare against FastEM here because it does not achieve comparable accuracy under any tested order setting in this case.

% Fig.~\ref{fig:timestep} (right) shows that, for the same number of time steps (200), the simulation time of FastEM increases noticeably as its model order is varied, whereas \textit{ExtRaKrylovEM} remains significantly faster. Because FastEM typically requires orders ranging from 15 to over 100,
% depending on the wire geometry and the desired accuracy, its runtime can vary substantially from design to design. Consequently, the speedup achieved by \textit{ExtRaKrylovEM} over FastEM also depends on the required FastEM order. 

% Fig.~\ref{fig:stress_arm_tree0} presents example stress maps generated by \textit{ExtRaKrylovEM} and \textit{EiRaKrylovEM} using the baseline configuration \{$\eta_{\mathrm{nuc}} = 1$ and $\eta_{\mathrm{post}} = 1$, $q = 6$\}, for both methods on the ARM tree circuit with 33 segments. The results from the FDM-based solver~\cite{SunYu:TDMR'20} are also shown for comparison at the top. Three time points are illustrated: around the nucleation instant (left), the mid–post-void stage (center), and the end of the post-void phase (right). The three set of figures actually are almost identical visually. Actually {ExtRaKrylovEM}  achieving a nucleation-time error of only
% 0.25\% and a resistance-change error of 0.00\% (bottom figure). 

% ===================================== Revised language writing ================================
% ===================================== Revised language writing ================================
% ===================================== Revised language writing ================================
% \include{figtex/Stress_comparison}
\begin{figure}[htbp]
    \centering
    \begin{subfigure}[b]{0.8\linewidth}
        \centering
        \includegraphics[width=\linewidth]{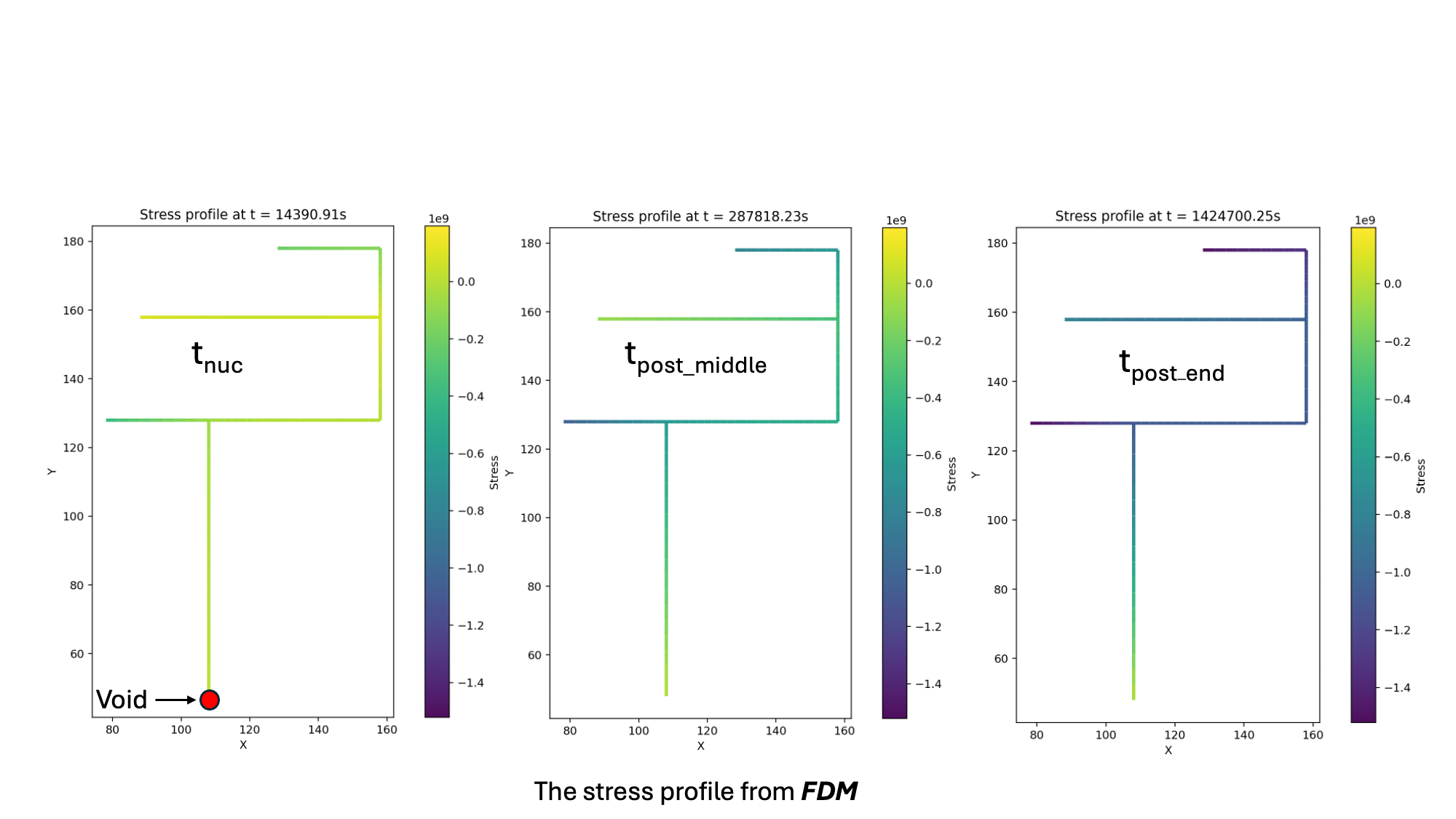} % Add figure name here
        \caption{Post void stress distribution obtained from FDM~\cite{SunYu:TDMR'20} under different time points.}
        \label{fig:wire0_10seg_fdm}
    \end{subfigure}
    \hfill
    \begin{subfigure}[b]{0.8\linewidth}
        \centering
        \includegraphics[width=\linewidth]{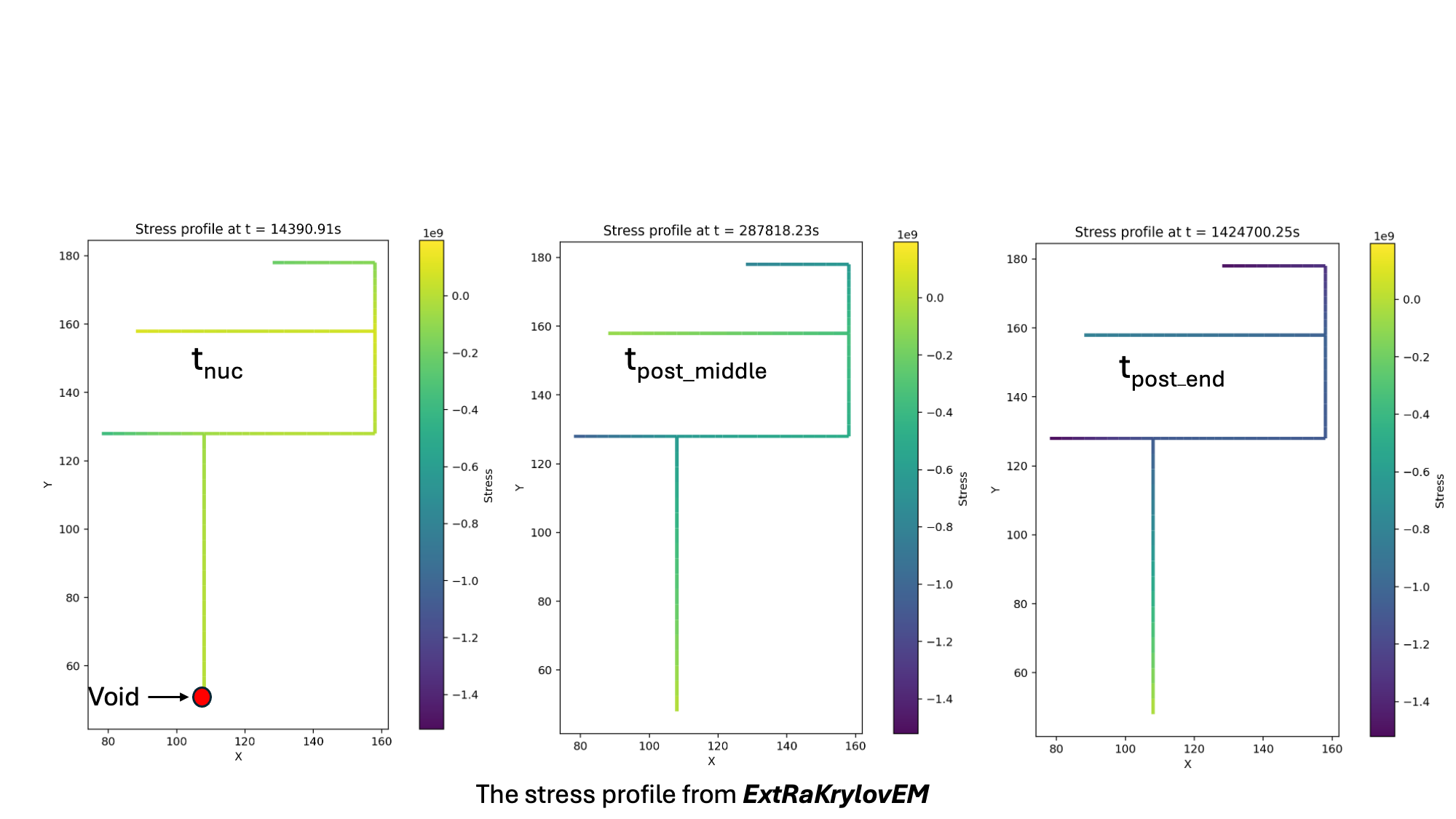} % Add figure name here
        \caption{Post void stress distribution obtained from the proposed \textit{ExtRaKrylovEM} under different time points.}
        \label{fig:wire0_10seg_rakrylov}
    \end{subfigure}
    \hfill
    \begin{subfigure}[b]{0.8\linewidth}
        \centering
        \includegraphics[width=\linewidth]{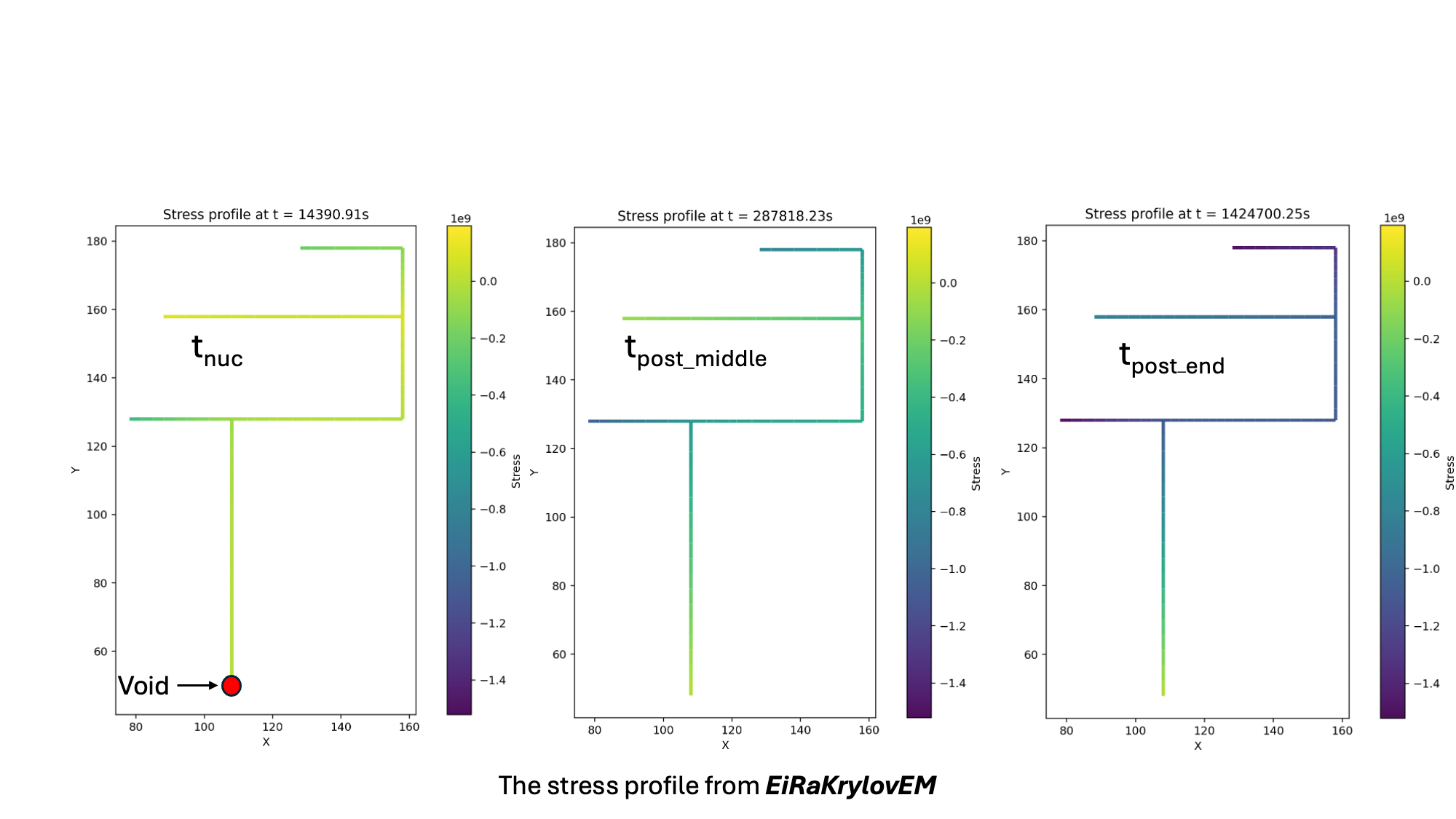} % Add figure name here
        \caption{Post void stress distribution obtained from the proposed \textit{EiRaKrylovEM} under different time points.}
        \label{fig:arm_tree0_ei}
    \end{subfigure}
    % \vspace{-5pt}
    \caption{\small Accuracy comparison with the finite difference method~\cite{SunYu:TDMR'20} at different times. Red dot indicates the nucleated cathode node}
    \label{fig:stress_wire0_10seg}
    \vspace{-10pt}
\end{figure}

Fig.~\ref{fig:stress_wire0_10seg} illustrates representative stress maps generated
by \textit{ExtRaKrylovEM} and \textit{EiRaKrylovEM} under the baseline configuration
$\{\eta_{\mathrm{nuc}}, \eta_{\mathrm{post}}, q\} = \{1,\, 1,\, 6\}$ with $\sigma_{\mathrm{crit}} = 10^8$~Pa on a synthesized 10-segment wire.
The corresponding results obtained from the FDM-based solver~\cite{SunYu:TDMR'20}
are shown at the top for reference.
Stress distributions at three representative time points are presented: around the nucleation instant (left), the middle of the post-void phase (center), and the end of
the post-void phase (right).
As shown in the figure, the stress maps produced by the three methods are
visually indistinguishable across all stages.
Quantitatively, \textit{ExtRaKrylovEM} yields a nucleation-time error of only
0.4\% and a resistance-change error of 0.00\%, as shown in the bottom panel.
Similar accuracy is achieved by \textit{EiRaKrylovEM}.

\section{Conclusion}
\label{sec:Conclusion}
% In this paper, we introduce fast variational analysis for electromigration (EM) damage analysis for VLSI interconnects.  We proposed a novel extended rational Krylov subspace method {\it ExtRaKrylovEM}, which allows time shifting  of the krylov subspace to the problems of interests such as nucleation time and post-void steady state time to achieve high fidelity reduced order models. As a result, it can just use a few order (4-6) to achieve near zero accuracy loss for all the benchmark wires we tested, which is in constrast with 15-50 orders required for transitional Krylov subspace method. 
% Experimental results on synthesized structures and an ARM chip design demonstrate that {\it ExtRaKrylovEM} achieves near-zero error in nucleation time and resistance
% change using only 4-6 order, while delivering an order-of-magnitude speedup over finite-difference solutions. Compared with standard extended Krylov methods—which typically require orders of 15-60 the proposed technique provides roughly $2\times$ additional speedup with even better accuracy.
% Furthermore, {\it ExtRaKrylovEM} enables highly efficient, high-fidelity MC analysis with almost no accuracy loss and up to two orders of magnitude speedup over finite-difference–based MC workflows.
In this paper, we presented two fast EM stress analysis techniques based on {\it rational Krylov} subspace reduction for electromigration (EM) reliability analysis of VLSI interconnects. Unlike traditional Krylov subspace methods, which can be viewed as expansions at infinite time (frequency 0), the rational Krylov subspace allows expansion at specific time constants aligned with application metrics such as nucleation time. We explored two simulation frameworks: {\it ExtRaKrylovEM}, based on the extended rational Krylov subspace method in the frequency domain, and {\it EiRaKrylovEM}, based on rational Krylov exponential integration in the time domain. We demonstrated that the accuracy of both methods is sensitive to the choice of expansion point, or equivalently, shift time, and showed that effective shift times are typically close to times of interest such as the nucleation time or the steady-state time. To further improve fidelity, we developed a coordinate-descent optimization method to identify the optimal reduction orders and shift times for both the nucleation and post-void phases.
Experimental results on synthesized structures and industrial benchmarks demonstrate that {\it ExtRaKrylovEM} and {\it EiRaKrylovEM} achieve orders-of-magnitude improvements in both efficiency and accuracy. Specifically, using only 4--6 reduction orders, our methods deliver sub-0.1\% error in nucleation-time and resistance-change predictions while providing 20--500$\times$ speedup over finite-difference solutions, depending on the number of simulation steps. In stark contrast, standard extended Krylov methods require 50+ orders yet still exhibit 10--20\% nucleation-time errors, rendering them impractical for EM-aware optimization and stochastic EM analysis.
%\input{doc/appdx}

% \bibliographystyle{ieee_fullname}
% \bibliographystyle{plain}
%\balance
% \bibliographystyle{IEEEtran}
% \bibliographystyle{ACM-Reference-Format}
\bibliographystyle{unsrt}
%\bibliography{ref/EM_basics, ref/mscad, ref/neural_network,ref/mscad_pub,ref/reduction,ref/reliability}
\bibliography{./ref/EM_basics, ../../bib/reliability_papers,../../bib/neural_network,../../bib/mscad_pub,../../bib/reduction,../../bib/reliability,../../bib/simulation}

\end{document}